\documentstyle[12pt,epsfig]{article}
\renewcommand{\theequation}{\thesection.\arabic{equation}}
\newcommand{\gsim}{\raisebox{-0.07cm}{$\:\stackrel{>}{{\scriptstyle
 \sim}}\: $} }
\newcommand{\lsim}{\raisebox{-0.07cm}{$\:\stackrel{<}{{\scriptstyle
 \sim}}\: $} }
\begin{document}
\setlength{\parskip}{0.5cm}
\setlength{\baselineskip}{0.6cm}
\begin{titlepage}

\noindent
{\tt hep-ph/9907472} \hfill INLO-PUB 14/99 \\ 
\hspace*{\fill} July 1999 \\
\vspace{1.5cm}
\begin{center}
\Large
{\bf NNLO Evolution of Deep-Inelastic} \\
\vspace{0.15cm}
{\bf Structure Functions: the Non-Singlet Case} \\
\vspace{2.2cm}
\large
W.L. van Neerven and A. Vogt \\
\vspace{0.8cm}
\normalsize
{\it Instituut-Lorentz, University of Leiden \\
\vspace{0.1cm}
P.O. Box 9506, 2300 RA Leiden, The Netherlands} \\
\vspace{4.5cm}
{\bf Abstract}
\vspace{-0.3cm}
\end{center}
We study the next-to-next-to-leading order (NNLO) evolution of flavour 
non-singlet quark densities and structure functions in massless 
perturbative QCD.
Present information on the corresponding three-loop splitting functions 
is used to derive parame\-trizations of these quantities, including 
Bjorken-$x$ dependent estimates of their residual uncertainties. 
Compact expressions are also provided for the exactly known, but rather 
involved two-loop coefficient functions.
The size of the NNLO corrections and their effect on the stability 
under variations of the renormalization scale are investigated. The 
residual uncertainty of the three-loop splitting functions does not 
lead to appreciable effects for $x > 10^{-2}$. Inclusion of the NNLO 
contributions reduces the main theoretical uncertainty of $\alpha_s$ 
determinations from non-singlet scaling violations by more than a 
factor of two.
\vfill
\end{titlepage}
%
%
\section{Introduction}
%
%
More than thirty years after the pioneering experiments at SLAC~\cite
{SLAC}, structure functions in deep-inelastic lepton-hadron scattering 
(DIS) remain among the most important probes of perturbative QCD and of 
the partonic structure of hadrons. Indeed, experiments have proceeded 
towards very high accuracy and a greatly extended kinematic coverage 
during the past two decades~\cite{exp}. Moreover, the forthcoming
luminosity upgrade of the electron-proton collider HERA at DESY will 
allow for accurate measurements up to very high resolution scales 
$Q^2 \simeq 10^4$ GeV$^2$, thus considerably increasing the lever arm 
for precise determination of the scaling violations, i.e., the 
$Q^2$-dependence, of the structure functions. An accurate knowledge of
the parton densities will also be indispensable for interpreting many
results at the future Large Hadron Collider at CERN.
 
Given the non-perturbative Bjorken-$x$ dependence of the structure
functions at one scale, the scaling violations can be calculated in 
the QCD-improved parton model in terms of a power expansion in the 
strong coupling constant $\alpha_s$. The next-to-leading order (NLO) 
ingredients for such analyses are available since 1980 for unpolarized 
structure functions in massless perturbative QCD~\cite{FP82}. Yet the 
corresponding results for the next-to-next-to-leading order (NNLO) are 
not complete at present, due to the enormous complexity of the required
loop calculations. Of the components entering the NNLO description, the 
three-loop $\beta$-function (governing the scale dependence of the 
strong coupling constant) \cite{beta2} and the two-loop contributions 
to the coefficient functions (connecting the structure functions to the 
parton densities) \cite{c2Sa,ZvN1,ZvN2} have been derived. However, 
only partial results have been obtained so far for the three-loop terms 
of the splitting functions (governing the scale-dependence of the quark 
and gluon densities), most notably the lowest even-integer Mellin 
moments of those combinations relevant to unpolarized electromagnetic 
deep-inelastic scattering~\cite{spfm}. 

Standard global analyses of deep-inelastic scattering and related 
processes, like the Drell--Yan process for which two-loop coefficient
functions have also been calculated \cite{c2DY}, have thus been 
restricted to NLO up to now \cite{MRS,CTEQ,GRV}. This level of accuracy
is however not sufficient to make full use of present and forthcoming 
data, as the theoretical uncertainties of the NLO results, for instance 
on the strong coupling constant, already now tend to exceed the 
corresponding experimental errors. Therefore first approximate NNLO 
analyses have been performed recently of data on neutrino-nucleon 
\cite{KPS} and electron$\,$(muon)-proton \cite{SY99} DIS structure
functions, directly using the results of refs.~\cite{spfm} via integer 
Mellin-$N$ techniques. However, these techniques lack some flexibility, 
e.g., they cannot incorporate additional information on the 
$x$-dependence of the two-loop coefficient functions \cite{ZvN1,ZvN2} 
and of the three-loop splitting functions 
\cite{Gra1,Gra2,CH94,BVns,BNRV}. 
Hence we pursue an alternative approach which allows for incorporating
the NNLO corrections into programs using standard $x$-space 
\cite{MRS,CTEQ} or equivalent complex-$N$ techniques~\cite{GRV,cmom}.
Its most important ingredients are compact approximate $x$-space 
expressions for the three-loop splitting functions including 
quantitative estimates of their present uncertainty. In the present 
article, we deal with the important flavour non-singlet case. The 
flavour-singlet quantities will be discussed in a subsequent 
publication.

This paper is organized as follows: In Sect.~2 we recall the general 
formalism for the scale dependence (`evolution') of non-singlet quark 
densities and structure functions in massless perturbative QCD. The 
$\alpha_s$-expansions are explicitly given up to NNLO for arbitrary
choices of the renormalization and mass-factorization scales. In 
Sect.~3 we present accurate, compact parametrizations of the exactly
known \cite{c2Sa,ZvN1,ZvN2}, but rather involved $x$-dependence of 
the two-loop coefficient functions. In Sect.~4 we employ the present
constraints \cite{spfm,Gra1,BVns} on the three-loop non-singlet 
splitting functions for deriving approximate expressions for their
$x$-dependence. The remaining uncertainties are quantified. All these 
results are put together in Sect.~5 to study the impact of the NNLO 
contributions on the evolution of the various non-singlet parton 
densities and structure functions. Here we also discuss the 
implications on determinations of $\alpha_s$ from DIS structure
functions. Finally our findings are summarized in Sect.~6. Mellin-$N$ 
space expressions for our parametrizations of the two-loop coefficient
functions of Sect.~3 can be found in the appendix.
%
%
%
\newpage
\section{The general formalism}
%
%
We set up our notations by recalling the NNLO evolution equations for 
non-singlet parton densities and structure functions. The number
distributions of quarks and antiquarks in a hadron are denoted by 
$q_i(x,\mu_f^2,\mu_r^2)$ and  $\bar{q}_i(x,\mu_f^2,\mu_r^2)$, 
respectively, where $x$ represents the fraction of the hadron's 
momentum carried by the parton. $\mu_r$ and $\mu_f$ stand for the 
renormalization and mass-factorization scales, and the subscript $i$ 
indicates the flavour of the (anti-)quark, with $i = 1, \ldots , N_f$ 
for $N_f$ flavours of effectively massless quarks.

The scale dependence of non-singlet combinations of these quark 
densities is governed by the (anti-)quark (anti-)quark splitting 
functions. Suppressing the dependence on $x$, $\mu_r$ and $\mu_f$ for 
the moment, the general structure of these functions, constrained by 
charge conjugation invariance and flavour symmetry, is given by
\begin{eqnarray}
  {\cal P}_{q_{i}q_{k}} \: = \: {\cal P}_{\bar{q}_{i}\bar{q}_{k}} 
  & = & \delta_{ik} {\cal P}_{qq}^V + {\cal P}_{qq}^S \nonumber \\
  {\cal P}_{q_{i}\bar{q}_{k}} \: = \: {\cal P}_{\bar{q}_{i}q_{k}} 
  & = & \delta_{ik} {\cal P}_{q\bar{q}}^V + {\cal P}_{q\bar{q}}^S 
  \:\: .
\end{eqnarray}
In an expansion in powers of the strong coupling constant $\alpha_s$ 
the flavour-diagonal (`valence') quantity ${\cal P}_{qq}^V$ starts at
first order, while ${\cal P}_{q\bar{q}}^V$ and the flavour-independent 
(`sea') contributions ${\cal P}_{qq}^S$ and ${\cal P}_{q\bar{q}}^S$ are 
of order $\alpha_s^2$. A non-vanishing difference ${\cal P}_{qq}^S - 
{\cal P}_{q\bar{q}}^S$ occurs for the first time at third order. This 
general structure leads to three independently evolving types of 
non-singlet distributions: The evolution of the flavour asymmetries
\begin{equation}
  q_{{\rm NS},ik}^{\pm} = q_i \pm \bar{q}_i - (q_k \pm \bar{q}_k)
\end{equation}
and of linear combinations thereof, hereafter generically denoted by 
$q_{\rm NS}^{\pm}$, is governed by 
\begin{equation}
\label{ppm} 
  {\cal P}_{\rm NS}^{\pm} = {\cal P}_{qq}^V  
  \pm {\cal P}_{q\bar{q}}^V \:\: .
\end{equation}
The sum of the valence distributions of all flavours, 
\begin{equation}
  q_{\rm NS}^V = \sum_{r=1}^{N_f} (q_r - \bar{q}_r) \:\: ,
\end{equation}
evolves with
\begin{equation}
\label{pval}
 {\cal P}_{\rm NS}^V = {\cal P}_{qq}^V - {\cal P}_{q\bar{q}}^V
 + N_f ({\cal P}_{qq}^S - {\cal P}_{q\bar{q}}^S) \:\: .
\end{equation}
The first moments of ${\cal P}_{\rm NS}^-$ and ${\cal P}_{\rm NS}^V$ 
vanish,
\begin{equation}
\label{mom1}
  \int_0^1 \! dx \: x^{N-1}\, {\cal P}_{\rm NS}^- \: = \:  
  \int_0^1 \! dx \: x^{N-1}\, {\cal P}_{\rm NS}^V \: = \: 0
  \quad \mbox{ for } \quad  N = 1 \:\: , 
\end{equation}
since the first moments of $q_{\rm NS}^-$ and $q_{\rm NS}^V$ reflect 
conserved additive quantum numbers.

The difference ${\cal P}_{qq}^S - {\cal P}_{q\bar{q}}^S$ is unknown 
except for the first moment, which vanishes by virtue of Eqs.\ 
(\ref{ppm}), (\ref{pval}) and (\ref{mom1}). However, the size of the 
2-loop contributions to ${\cal P}_{q\bar{q}}^V$ and ${\cal P}_{qq}^S$ 
relative to the corresponding term of ${\cal P}_{qq}^V$ suggests that 
this difference is negligibly small at moderate and large $x$. Hence 
we shall use the approximation
\begin{equation}
  {\cal P}_{\rm NS}^V \: = \: {\cal P}_{\rm NS}^- 
\end{equation}
for the rest of this article, i.e., we henceforth treat $q_{\rm NS}^V$ 
as a `--'-quantity. 

Restoring the dependence on the fractional momentum $x$ and the 
renormalization and mass-factorization scales $\mu_r$ and $\mu_f$, our
evolution equations thus read
\begin{equation}
\label{evol}
  \frac{d}{d \ln \mu_f^2} \, q_{\rm NS}^{\pm}(x,\mu_f^2, \mu_r^2) 
  \: = \:
  \Big[ {\cal P}_{\rm NS}^{\pm} \bigg( \alpha_s(\mu_r^2), \frac{\mu_f^2}
  {\mu_r^2} \bigg) \otimes q_{\rm NS}^{\pm}(\mu_f^2,\mu_r^2) \Big](x) 
  \:\: .
\end{equation}
Here $\otimes$ stands for the Mellin convolution in the momentum 
variable,
\begin{equation}
  [ a \otimes b ](x) \: = \: \int_x^1 \! \frac{dy}{y} \: a(y)\, 
  b\bigg(\frac{x}{y}\bigg) \:\: .
\end{equation}
The expansion of ${\cal P}_{\rm NS}^{\pm}$ up to the third
order (NNLO) in $a_s \equiv\alpha_s /(4\pi)$ takes the form 
\begin{eqnarray}
  {\cal P}_{\rm NS}^{\pm}\Big( x, \alpha_s(\mu_r^2), 
    \frac{\mu_f^2}{\mu_r^2} \Big) 
  & = & \quad  
      a_s(\mu_r^2) \, P_{\rm NS}^{(0)}(x) \nonumber \\ 
  & & \mbox{}\!\! 
    + a_s^2(\mu_r^2) \, \bigg( P_{\rm NS}^{(1)\pm}(x) - \beta_0 
      P_{\rm NS}^{(0)}(x) \ln \frac{\mu_f^2}{\mu_r^2} \bigg) \:  \\
  & & \mbox{}\!\!
    + a_s^3(\mu_r^2) \, \bigg( P_{\rm NS}^{(2)\pm}(x) 
    - \bigg\{ \beta_1 P_{\rm NS}^{(0)}(x) + 2\beta_0 P_{\rm NS}^{(1)\pm}
   (x) \bigg\} \ln \frac{\mu_f^2}{\mu_r^2} \nonumber \\
  & & \mbox{} \quad\quad\quad\quad
    + \beta_0^2 P_{\rm NS}^{(0)}(x) \ln^2 \frac{\mu_f^2}{\mu_r^2}\, 
      \bigg) + \ldots \nonumber \:\: .
\end{eqnarray}
The one- and two-loop functions $P_{\rm NS}^{(0)}(x)$ and 
$P_{\rm NS}^{(1)\pm}(x)$ are known for a long time \cite{FP82};
the 3-loop quantities $P_{\rm NS}^{(1)\pm}(x)$ are the subject of 
Sect.~4. The relevant coefficients of the QCD $\beta$-function,
\begin{equation}
  \frac{d a_s}{d \ln \mu_r^2} \: = \: \beta(a_s) \: = \: 
  - \sum_{l=0} a_s^{l+2} \beta_l \:\: ,
\end{equation} 
are given by \cite{FP82,beta2}
\begin{eqnarray}
  \beta_0 & = &   11 - \frac{2}{3}  N_f  \nonumber \\
  \beta_1 & = &  102 - \frac{38}{3} N_f  \\
  \beta_2 & = &  \frac{2857}{2} - \frac{5033}{18} N_f +
                 \frac{325}{54} N_f^2  \:\: .   \nonumber
\end{eqnarray}
The first two coefficients $\beta_0$ and $\beta_1$ are scheme 
independent in massless QCD; the result given for $\beta_2$ refers to 
the $\overline{\mbox{MS}}$ renormalization scheme employed throughout 
this paper.

The non-singlet structure functions $F_{a,{\rm NS}}^{\pm}$, $a = 1, 2, 
3$, are in Bjorken-$x$ space obtained by convoluting the solution of 
Eq.~(\ref{evol}) with the corresponding coefficient functions:
\begin{equation}
 \eta_a F_{a,\rm NS}^{\pm}(x,Q^2)\: = \:\Big[ {\cal C}_{a,\rm NS}^{\pm}
  \Big( \alpha_s(\mu_r^2), \frac{Q^2}{\mu_f^2}, \frac{\mu_f^2}{\mu_r^2} 
  \Big) \otimes q_{\rm NS}^{\pm}(\mu_f^2, \mu_r^2) \Big] (x)
\end{equation}
with $\eta_1 = 2$, $\eta_2 = 1/x$, $\eta_3 = 1$, and
\begin{eqnarray}
\label{cnnlo}
   \lefteqn{ {\cal C}_{a,\rm NS}^{\pm}\Big( x, \alpha_s(\mu_r^2),
   \frac{Q^2}{\mu_f^2}, \frac{\mu_f^2}{\mu_r^2} \Big) \: = } 
   \nonumber \\
 & & \quad\! 
   \delta (1-x) \: + \: a_s(\mu_r^2) \bigg( c_{a,\rm NS}^{(1)}(x) + 
   P_{\rm NS}^{(0)}(x) \ln \frac{Q^2}{\mu_f^2} \bigg) \\
 & & \mbox{}\!\!\!\! 
   + a_s^2(\mu_r^2) \: \bigg( c_{a,\rm NS}^{(2)\pm }(x) + \bigg\{
   P_{\rm NS}^{(1)\pm}(x) + \Big[ P_{\rm NS}^{(0)} \otimes 
   c_{a,\rm NS}^{(1)} \Big] (x) \bigg\} \ln \frac{Q^2}{\mu_f^2} 
   - \beta_0 c_{a,{\rm NS}}^{(1)}(x) \ln \frac{Q^2}{\mu_r^2} 
   \nonumber \\
 & & \mbox{} \quad\quad\quad
   + \frac{1}{2} \bigg\{ \Big[ P_{\rm NS}^{(0)}\otimes P_{\rm NS}^{(0)}
   \Big] (x) - \beta_0  P_{\rm NS}^{(0)}(x) \bigg\}
   \ln^2 \frac{Q^2}{\mu_f^2} - \beta_0 P_{\rm NS}^{(0)}(x) 
   \ln \frac{Q^2}{\mu_f^2} \ln \frac{\mu_f^2}{\mu_r^2} 
   \bigg) + \ldots \:\: .
   \nonumber
\end{eqnarray}
Here an overall electroweak charge factor has been absorbed into 
$q_{\rm NS}^{\pm}$. The first-order coefficients $c_{a,{\rm NS}}^{(1)}
(x)$ can be found in ref.~\cite{FP82}; the 2-loop quantities 
$c_{a,{\rm NS}}^{(2)\pm}(x)$ computed in refs.~\cite{c2Sa,ZvN1} are 
discussed in Sect.~3.

It is often convenient, especially in the non-singlet sector considered 
here, to express the scaling violations of the structure functions in 
terms of these structure functions themselves. The expansion 
coefficients of the corresponding kernels ${\cal K}_{a,\rm NS}^{\pm}$ 
in 
\begin{equation}
  \frac{d}{d \ln Q^2} \, F_{a,\rm NS}^{\pm}(x,Q^2)
  \: = \:
  \Big[ {\cal K}_{a,\rm NS}^{\pm} \bigg( \alpha_s(\mu_r^2), 
  \frac{Q^2}{\mu_r^2} \bigg) \otimes F_{a,\rm NS}^{\pm}(Q^2) 
  \Big] (x)
\end{equation}
are built up of factorization-scheme invariant combinations of the 
splitting functions $P_{\rm NS}^{(l)\pm}(x)$ and the coefficient 
functions $c_{a,{\rm NS}}^{(l)\pm}(x)$. Up to third order this
expansion reads 
\begin{eqnarray} 
   {\cal K}_{\rm NS}^{\pm} \bigg( x, \alpha_s(\mu_r^2), \frac{Q^2}
   {\mu_r^2} \bigg)
 \!\! & = & \!\!\!\quad
     a_s(\mu_r^2) \: P_{\rm NS}^{(0)}(x) \nonumber \\
 & & \mbox{}\!\!\!\!\!
   + a_s^2 (\mu_r^2) \: \bigg( P_{\rm NS}^{(1)\pm}(x) 
   - \beta_0 c_{a,{\rm NS}}^{(1)}(x) - \beta_0 P_{\rm NS}^{(0)}(x) 
   \ln \frac{Q^2}{\mu_r^2} \bigg) \\
 & & \mbox{}\!\!\!\!\!
   + a_s^3(\mu_r^2) \: \bigg( P_{\rm NS}^{(2)\pm}(x)  
   - 2 \beta_0 c_{a,{\rm NS}} ^{(2)\pm}(x)
   + \beta_0 \Big[ c_{a,{\rm NS}}^{(1)} \otimes c_{a,{\rm NS}}^{(1)} 
   \Big] (x) \nonumber \\
 & & \mbox{}\quad\quad\!
   - \beta_1 c_{a,{\rm NS}}^{(1)}(x)
   + \beta_0^2 P_{\rm NS}^{(0)}(x) \ln^2 \frac{Q^2}{\mu_r^2} 
   \nonumber \\
 & & \mbox{}\quad\quad\!
   - \bigg\{ 2 \beta_0 \{ P_{\rm NS}^{(1)\pm}(x) - \beta_0 
   c_{a,{\rm NS}}^{(1)}(x) \}
   + \beta_1 P_{\rm NS}^{(0)}(x) \bigg\} \ln \frac{Q^2}{\mu_r^2} \bigg)
   + \ldots \:\: . \nonumber
\end{eqnarray}
This approach removes the dependence of the finite-order 
predictions on the factorization scheme and the scale $\mu_f$, thus
allowing for an easier control of the theoretical uncertainties. 
%
%
\section{The 2-loop non-singlet coefficient functions}
%
\setcounter{equation}{0}
The $O(\alpha_s^2)$ contributions ${\cal C}^{(2)}_{a}$ to the 
coefficient functions for the structure functions $F_2$, $F_L = 
F_2 - 2x F_1$ and $xF_3$ were calculated some time ago in refs.\ 
\cite{c2Sa,ZvN1,ZvN2}. The resulting expressions are rather lengthy and 
involve higher transcendental functions. Hence it is convenient to 
employ more compact, if approximate, parametrizations of these 
quantities. This holds in particular if one uses the moment-space 
technique \cite{cmom}, which requires the analytic continuation of 
all ingredients to complex Mellin-$N$. The reader is referred to 
refs.~\cite{BlKu} for a more rigorous approach to the moment-space 
expressions for ${\cal C}^{(2)}_{a}$.
Those parts of the coefficient functions arising from $\mu_r \neq Q$
and $\mu_f \neq \mu_r$ in Eq.~(\ref{cnnlo}) are simple convolutions  
of the well-known lower-order anomalous dimensions and Wilson 
coefficients. The same applies to the terms induced by usual scheme 
transformations, e.g., that from the $\overline{\mbox{MS}}$ to the DIS 
factorization scheme. For explicit expressions see refs.~\cite{ZvN2}. 
Hence the parametrizations can be restricted to the 
$\overline{\mbox{MS}}$ scheme, and to $\ln (\mu_r^2/Q^2) = \ln 
(\mu_r^2/\mu_f^2) = 0$.

Our procedure for deriving compact approximate expressions for 
$c^{(2)}_{a, \rm NS}(x)$ is as follows:
We keep the +-distribution parts, defined by 
\begin{equation} 
  \int_0^1 \! dx \, a(x)_+ b(x) \: = \: \int_0^1 \! dx \, a(x)
  \,\{ b(x) - b(1) \} \:\: ,
\end{equation}
exactly (up to a truncation of the numerical coefficients). The 
integrable $x\! <\! 1$ terms are fitted to the exact results for 
$10^{-6} \leq x \leq 1\! -\! 10^{-6}$. Finally the coefficients of 
$\delta (1-x)$ are slightly adjusted from their exact values using the 
lowest integer moments. The resulting parametrizations deviate from the 
exact results by no more than a few permille. This holds for the 
$c^{(2)}_{a,\rm NS}(x\! <\! 1)$ themselves as well as for the 
convolutions with typical hadronic $x$-shapes. The adjustment of the 
$\delta (1-x)$ pieces is important for the latter agreement.	

The non-singlet coefficient function entering the electromagnetic
$F_2$ can be written as 
\begin{eqnarray}
  c_{2,{\rm NS}}^{(2)+}(x) 
   &=& \bigg( \frac{1}{1-x} \,\Big[ 14.2222\, L_1^3  
       - 61.3333\, L_1^2 - 31.105\, L_1  + 188.64 \Big] \bigg)_+ 
       \nonumber \\
   & & \mbox{}- 17.19\, L_1^3 + 71.08\, L_1^2 - 660.7\, L_1   
       + L_1 L_0 (- 174.8\, L_1 +  95.09\, L_0) 
       \nonumber \\
   & & \mbox{}- 2.835\, L_0^3 - 17.08\, L_0^2 + 5.986\, L_0   
       - 1008\, x - 69.59 - 338.046\, \delta (1-x) 
       \nonumber \\
   & & \hspace*{-7mm}\mbox{} + N_f\: \Bigg\{ \bigg( \frac{1}{1-x} 
       \Big[ 1.77778 \, L_1^2 - 8.5926\, L_1 + 6.3489 \Big] \bigg)_+ 
       \nonumber \\
   & & \mbox{}- 1.707\, L_1^2 + 22.95\, L_1 + L_1 L_0 (3.036\, L_0 
       + 17.97) 
       \\
   & & \mbox{}+ 2.244\, L_0^2 + 5.770\, L_0 - 37.91\, x - 5.691\, 
       + 46.8405\, \delta (1-x) \Bigg\} \nonumber
\end{eqnarray}    
%
with 
$$
  L_1 \:\equiv\: \ln (1-x) \:\: , \:\:\:\: L_0 \:\equiv\: \ln x \:\: .
$$
For $c_{2,{\rm NS}}^{(2)-}(x)$, relevant for the charged-current case,
the second and third line of this expression have to be replaced by
\begin{eqnarray}
   & & \mbox{}- 17.19\, L_1^3 + 71.08\, L_1^2 - 663.0\, L_1   
     + L_1 L_0 (- 192.4\, L_1 + 80.41\, L_0) \\
   & & \mbox{}- 3.748\, L_0^3 - 19.56\, L_0^2 - 1.235\, L_0   
       - 1010\, x - 84.18 - 337.994\, \delta (1-x) \:\: . \nonumber
\end{eqnarray} 
The corresponding parametrizations for $F_L$ read 
\begin{eqnarray}
 c_{L,{\rm NS}}^{(2)+}(x) 
  &\! = \! & 13.62\, L_1^2 - 55.79\, L_1 - 150.5\, L_1 L_0 
    + (26.56\, x - 0.031) L_0^2 - 14.85\, L_0 \quad\quad\quad \\     
  & & \mbox{}\!\!\!\!\! + 97.48\, x - 40.41 - 0.164\, \delta (1-x)
    + \frac{16}{27}\, N_f\: \Big\{ 6\, xL_1 - 12\, xL_0 - 25\, x 
    + 6 \Big\} \nonumber
\end{eqnarray}
and 
\begin{eqnarray}
  c_{L,{\rm NS}}^{(2)-}(x)
   &\! =\! & 13.30\, L_1^2 - 59.12\, L_1 - 141.7\, L_1 L_0
     + (23.29\, x - 0.043) L_0^2 - 22.21\, L_0 \quad\quad\quad \\
   & & \mbox{}\!\!\!\!\! + 100.8\, x - 52.27 - 0.150\, \delta (1-x)
     + \frac{16}{27}\, N_f\: \Big\{ 6\, xL_1 - 12\, xL_0 - 25\, x 
     + 6 \Big\} \nonumber \:\: . 
\end{eqnarray}
For the $N_f$ parts, which are identical in Eqs.~(3.4) and (3.5), we 
have taken the exact expression from ref.~\cite{c2Sa}.
 
Also for the charged-current non-singlet structure function $F_3$ there
are two combinations which differ at $O(\alpha_s^2)$. The first one, 
entering $F_3^{\nu N} + F_3^{\bar{\nu} N}$, can be written as
\begin{eqnarray}
  c_{3,\rm NS}^{(2)-}(x) 
   &=& \bigg( \frac{1}{1-x} \,\Big[ 14.2222\, L_1^3  
       - 61.3333\, L_1^2 - 31.105\, L_1  + 188.64 \Big] \bigg)_+ 
       \nonumber \\
   & & \mbox{}- 15.20\, L_1^3 + 94.61\, L_1^2 - 409.6\, L_1   
       - 147.9\, L_1^2 L_0 
       \nonumber \\
   & & \mbox{}- 3.922\, L_0^3 - 33.31\, L_0^2 - 67.60\, L_0   
       - 576.8\, x - 206.1 - 338.635\, \delta (1-x) 
       \nonumber \\
   & & \hspace*{-7mm}\mbox{} + N_f\: \Bigg\{ \bigg( \frac{1}{1-x} 
       \Big[ 1.77778 \, L_1^2 - 8.5926\, L_1 + 6.3489 \Big] \bigg)_+ 
       \nonumber \\
   & & \mbox{}+ 0.042\, L_1^3 - 0.808\, L_1^2 + 25.00\, L_1 
       + 9.684\, L_1 L_0 
       \\
   & & \mbox{}+ 2.207\, L_0^2 + 8.683\, L_0 - 14.97\, x - 6.337\, 
       + 46.857\, \delta (1-x) \Bigg\} \:\: . \nonumber
\end{eqnarray}   
For the other combination $c_{3,\rm NS}^{(2)+}(x)$, corresponding to
$F_3^{\nu N} - F_3^{\bar{\nu} N}$, the second and third line of this 
result have to be replaced by
\begin{eqnarray}
   & & \mbox{}- 15.20\, L_1^3 + 94.61\, L_1^2 - 396.1\, L_1   
       - 92.43\,  L_1^2 L_0 \\
   & & \mbox{}- 3.049\, L_0^3 - 30.14\, L_0^2 - 79.14\, L_0   
       - 467.2\, x - 242.9 - 338.683\, \delta (1-x) \:\: . \nonumber
\end{eqnarray} 

The complex Mellin moments of these results, $c_{a,\rm NS}^{(2)\pm}(N)$,
can be readily obtained. They do not involve special functions beyond 
the logarithmic derivatives of the $\Gamma$-function. The explicit 
expressions can be found in the appendix.
%
%
\section{The 3-loop non-singlet splitting functions}
%
\setcounter{equation}{0}
Only partial results are presently available for the $O(\alpha_s^3)$ 
terms $P^{(2)}(x)$ of the splitting functions. In the non-singlet
sector, the current information comprises
\begin{itemize}
\item the lowest five even-integer moments of $P^{(2)+}_{\rm NS}$ 
      calculated in refs.~\cite{spfm}, while for $P^{(2)-}_{\rm NS}$
      only the first moment (= 0 in $\overline{\rm MS}$) is known;
\item the complete $O(N_f^2)$ piece (identical for the `+' and `$-$' 
      combinations) determined via an all-order leading-$N_f$ approach
      in ref.~\cite{Gra1};
\item the most singular small-$x$ terms ($\sim \ln ^4 x$) of 
      $P_{\rm NS}^{(2)+}$ and $P_{\rm NS}^{(2)-}$ inferred in 
      ref.~\cite{BVns} from the leading small-$x$ resummation
      of the non-singlet evolution kernels \cite{KL83}. 
\end{itemize}
The 2-loop results $P^{(1)\pm}_{\rm NS}(x)$ \cite{CFP} and 
$c^{(2)\pm}_{a,\rm NS}(x)$ \cite{ZvN1} furthermore indicate that the 
difference $P^{(2)+}_{\rm NS}(x) - P^{(2)-}_{\rm NS}(x)$ is negligibly 
small at large $x$. Finally present knowledge complies with the 
conjecture \cite{GLY} that the splitting functions do not receive 
contributions of the form $[\ln^l (1-x)/(1-x)]_+$ with $l>0$ in the 
$\overline{\rm MS}$ factorization scheme, unlike the coefficient 
functions discussed above.

In what follows we employ this information for approximate 
reconstructions of 
\begin{equation}
\label{nf_dep}
  P_{\rm NS}^{(2)\pm}(x) = P_{0}^{(2)\pm}(x) + N_f  P_{1}^{(2)\pm}(x) 
  + N_f^2 P_{2}^{(2)}(x) \:\: .
\end{equation} 
Our approach is to fix the coefficients of suitably chosen basis 
functions by the above constraints. The spread of the result due to 
`reasonable' variations in the choice of those functions then provides 
a measure of the residual uncertainty. Specifically we employ the 
ansatz
\begin{equation}
\label{ansatz}
  P_{i}^{\pm}(x) =  \frac{A_{i,1}}{(1-x)_+} + A_{i,2}\,\delta (1-x)  
     + A_{i,3}^{\pm}\, f_1(x) + A_{i,4}^{\pm} f_m(x) 
     + A_{i,5}^{\pm} f_0(x) + f_{\rm as}^{\pm}(x)
\end{equation}
for the $N_f$-independent ($i=0$) and $N_F^1$ ($i=1$) terms in Eq.\ 
(\ref{nf_dep}).
Here $f_1$ and $f_0$ represent contributions which, while being 
integrable, peak at $x\!\rightarrow\! 1$ and $x\!\rightarrow\! 0$, 
respectively. $f_m$ stands for a part with a rather flat $x$-dependence.
As for the illustrations in ref.~\cite{spfm}, these contributions are
build up of powers of $\ln (1\! -\! x)$, $x$, and $\ln x$. Finally 
$f_{\rm as}$ allows to account for known leading small-$x$ terms. 
Equating the second to tenth even moments of Eq.~(\ref{ansatz}) to the 
results of ref.~\cite{spfm} yields five linear equations which can be 
solved for the coefficients $A_{i,j}^{+}$. The case of $P_{\rm NS}
^{(2)-}$ is treated afterwards by taking over $A_{i,1}$ and $A_{i,2}$ 
from the `+'-combinations, as already indicated in Eq.~(\ref{ansatz}), 
and adjusting the remaining coefficients as discussed below.

\begin{figure}[t]
\vspace*{-2mm}
\centerline{\epsfig{file=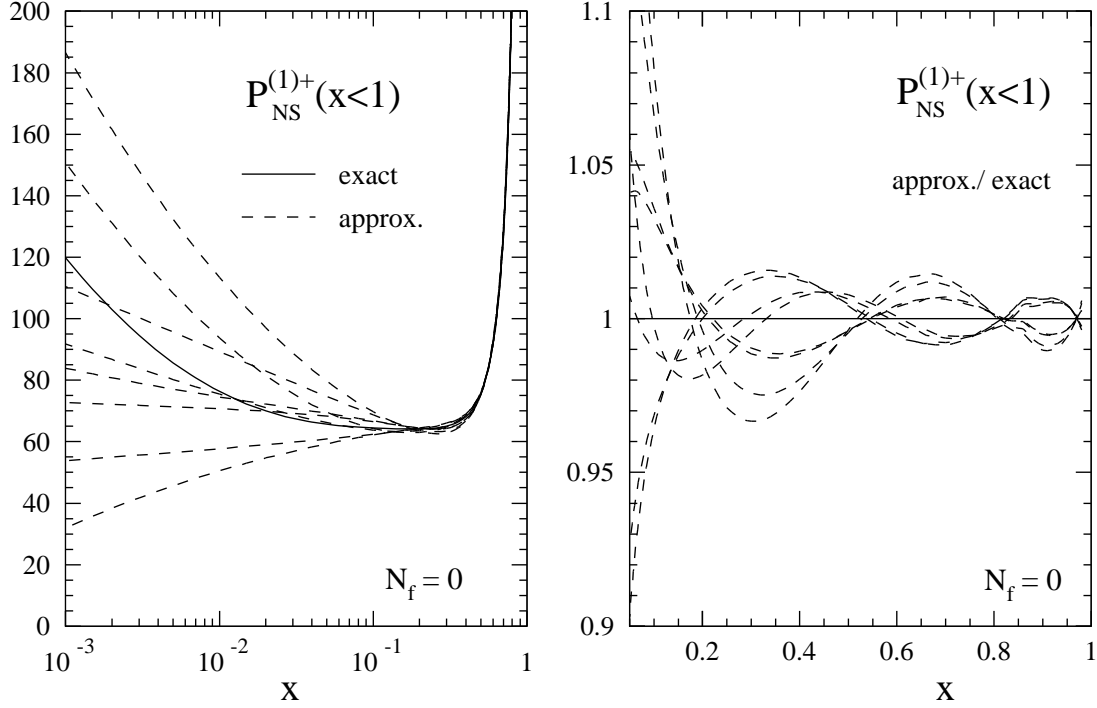,width=15cm,angle=0}}
\vspace*{-1mm}
\caption{Approximations for the $N_f$-independent part of $P^{(1)+}
 _{\rm NS}$, derived from the lowest even-integer moments by means 
 of Eqs.\ (\ref{ansatz}) and (\ref{pnlo}), compared to the exact 
 result.}
\end{figure}

Before addressing $P_{\rm NS}^{(2)\pm}(x)$ we demonstrate our procedure 
by applying it to a known result, the $N_f = 0$ part $P_0^{(1)+}$ of the 
NLO splitting function $P^{(1)+}_{\rm NS}(x)$ \cite{CFP}. In this case 
the leading small-$x$ contributions are $\ln x$ and $\ln^2 x$, while 
the integrable terms most peaked at large-$x$ read $x^2$ and 
$\ln (1\! -\! x) $. Disregarding small-$x$ constraints in this example, 
we thus choose
\begin{equation}
\label{pnlo}
  \begin{array}{ccccc}
  f_1(x) &=&  x^2  & \mbox{ or } & \ln (1-x)            \\[0.5mm]
  f_m(x) &=&   1   & \mbox{ or } &    x                 \\[0.5mm]
  f_0(x) &=& \ln x & \mbox{ or } & \ln^2 x              \\[2mm]
  f_{\rm as}(x) &\equiv & 0 & . \:\:\: &    
  \end{array}
\end{equation}
The resulting eight approximations are compared to the exact result in 
Fig.~1 for $x<1$. The latter curve runs inside the uncertainty band 
over the full $x$-range. The moments tightly constrain $P^{(1)+}_{0}(x)$
for $x \gsim 0.15$, the total spread in our approach being about 5\% at 
$x \simeq 0.3$.
The coefficients of the common leading large-$x$ 
terms are
\begin{eqnarray}
  A_{0,1}^{(1)} &=& 65.13 \ldots 68.74 \quad\quad (\mbox{exact } 66.47) 
  \nonumber \\
  A_{0,2}^{(1)} &=& 61.85 \ldots 79.64 \quad\quad (\mbox{exact } 69.00) 
  \:\: ,
\end{eqnarray}
and the first moments read
\begin{equation}
  P_{0}^{(1)+}(N\! =\! 1) 
  \: =\:  - 2.404 \ldots 0.400 \quad\quad (\mbox{exact } -1.127)
  \:\: .
\end{equation}
`Unreasonable' combinations in the sense of Eq.~(\ref{ansatz}), like 
$\ln (1 - x)$, $x^2$, and $1$ (i.e., no $f_0$) or $1$, $\ln x$, and 
$\ln^2 x$ (i.e., $f_1$ missing), can lead to considerably worse 
approximations. 

\begin{figure}[t]
\vspace*{-2mm}
\centerline{\epsfig{file=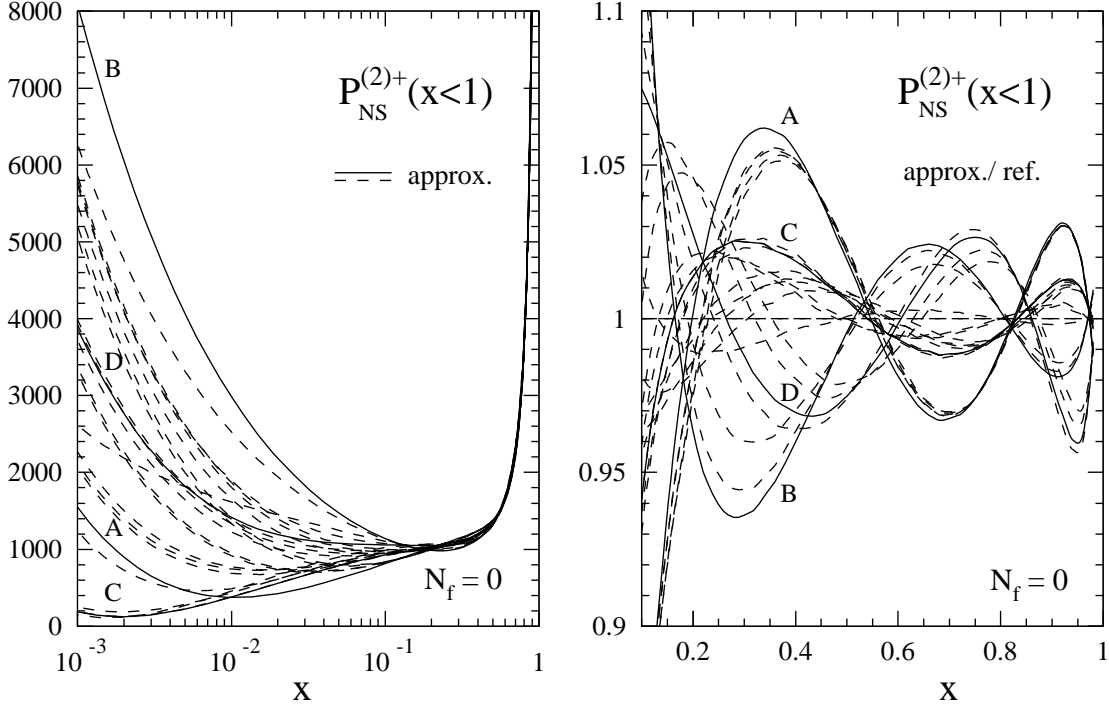,width=15cm,angle=0}}
\vspace*{-1mm}
\caption{Approximations for the $N_f$-independent part of $P^{(2)+}
 _{\rm NS}$, denoted by $P_0^{(2)+}$ in Eq.~(\ref{nf_dep}), as derived 
 from the five lowest even-integer moments by means of Eqs.\ 
 (\ref{ansatz}), (\ref{pnnlo0}) and (\ref{mo1}). The full lines 
 represent those functions selected for further consideration.}
\end{figure}
 
Now we turn to $P_0^{(2)+}(x)$. The additional loop or emission may,
besides adding two powers of $\ln x$, lead to two additional large-$x$ 
logarithms with respect to $P_0^{(1)+}(x)$ (the transition from 1-loop 
to 2-loop yields however only a term $\ln^1 (1-x)\, $). Hence we put
\begin{equation}
\label{pnnlo0}
  \begin{array}{ccccccccc}
  f_1(x) &=&  x^2  & \mbox{ or } & \ln (1-x) & \mbox{ or } 
           & \ln^2 (1-x) & \mbox{ or } & \ln^3 (1-x)          \\[0.5mm]
  f_m(x) &=&   1   & \mbox{ or } &    x    & &  & &           \\[0.5mm]
  f_0(x) &=& \ln x & \mbox{ or } & \ln^2 x & &  & &           \\[2mm]
  f_{\rm as}(x) &=& \multicolumn{4}{l} { {\displaystyle 
             \frac{2}{3}}\, C_F^3 \Big(\ln^4 x + \lambda \ln^3 x \Big) 
              \:\: . } & & &       
  \end{array}
\end{equation}
Besides $\lambda = 0$ we also include $\lambda = -4$ and $\lambda = 8$
for $f_0 = \ln x$. Subleading small-$x$ terms of this order of magnitude
are suggested by the expansion of $P_{\rm NS}^{(0)+}$ and $P_{\rm NS}
^{(1)+}$ in moment-space around $N=0$ \cite{BRV}. Thus we consider 32 
combinations, 8 of which are rejected as they fail to fulfill the
further ad hoc, but mild constraint 
\begin{equation}
\label{mo1}
     100\: P_{0}^{(1)+}(N\! =\! 1)\:\leq\: P_{0}^{(2)+}(N\! =\! 1)
     \:\leq\: -40\: P_{0}^{(1)+}(N\! =\! 1)
\end{equation}
on the perturbative expansion of the first moment. The $x\! >\! 1$ 
behaviour of the remaining 24 function is displayed in Fig.~2; their
$1/(1-x)_+$ coefficients span the range
\begin{equation}
     1138 \:\leq\: A_{0,1}^{(2)} \:\leq\: 1625 \:\: (1347) \:\: .
\end{equation}
The bracketed number applies if combinations with $f_1(x) = 
\ln^3 (1-x)$ are disregarded.

\begin{figure}[t]
\vspace*{-2mm}
\centerline{\epsfig{file=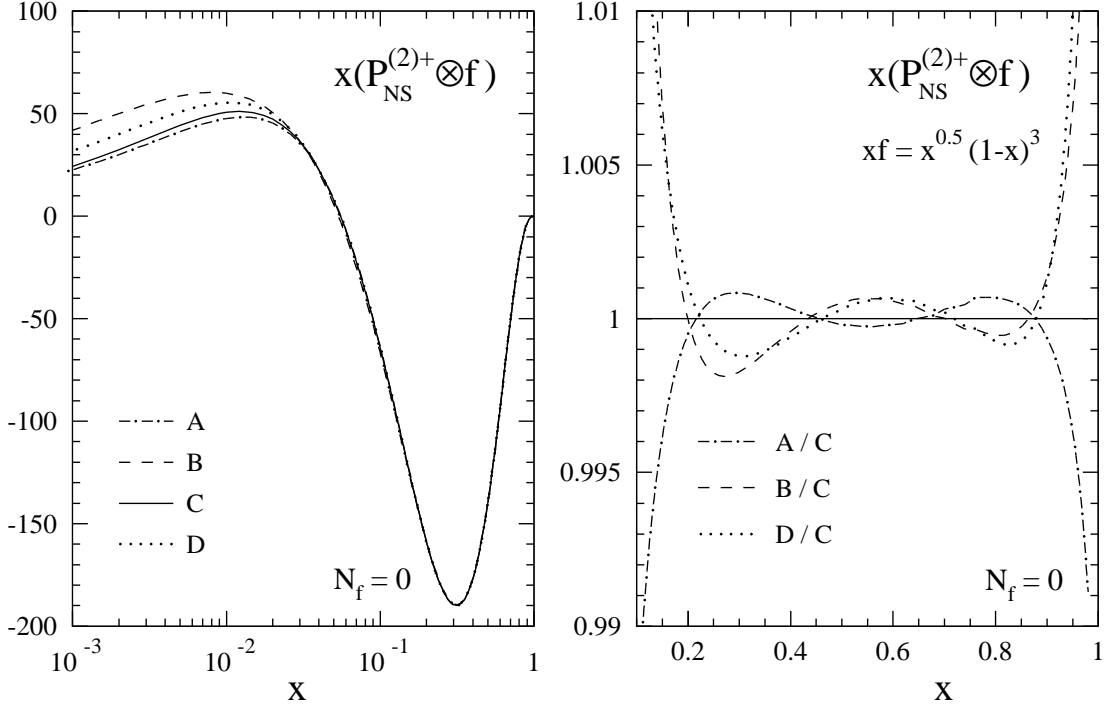,width=15cm,angle=0}}
\vspace*{-1mm}
\caption{The convolution of the approximations `A' -- `D' of 
 $P_0^{(2)+}$ selected in Fig.~2 with a shape typical of hadronic 
 non-singlet initial distributions.} 
\end{figure}

Due to the larger function pool of Eq.~(\ref{pnnlo0}), the large-$x$ 
uncertainty band of Fig.~2 is some factor of three wider than that for
$P_0^{(1)+}(x)$ in Fig.~1, reaching a total spread of about 15\% at 
$x \simeq 0.3$. Moreover $P_0^{(2)+}(x)$ is rather unconstrained 
at small $x \lsim 10^{-2}$ by present information, as the leading 
small-$x$ term \cite{BVns} does not dominate over less singular 
contributions at practically relevant values of $x$. 
However, physical quantities are only affected by the splitting 
functions via convolutions with smooth non-perturbative initial 
distributions which `wash out' the oscillating large-$x$ differences of 
Fig.~2 to a large extent. Furthermore the convolutions receive 
important contributions from the (well-constrained) large-$x$ region of 
$P_{\rm NS}(x)$ even at very small $x$. The above `bare' uncertainty is 
thus considerably reduced over the full $x$-range. This effect is 
illustrated in Fig.~3, where four representative approximate results 
for $P_0^{(2)+}$ are convoluted with a simple, but typical input shape. 
The total spread after this convolution is as small as 0.3\% for $0.2 
\lsim x \lsim 0.9 $, and becomes large only at $x \lsim 0.02 $. 

The uncertainty band of Fig.~3 is rather completely covered by the 
results `A' and `B'. Hence our final estimates for $P^{(2)+}_{0}(x)$ 
and its remaining uncertainty are given by
\begin{eqnarray}
\label{nsp0a}
  P^{(2)+}_{0,\, A}(x)\! &=& 
     1137.897\: \frac{1}{(1-x)_+} + 1099.754\: \delta (1-x) 
     - 2975.371\, x^2 \nonumber \\
  & & \mbox{} - 125.243 - 64.105\: \ln^2 x + 1.580\:\ln^4 x \\
  P^{(2)+}_{0,\, B}(x)\! &=&   
     1347.207\: \frac{1}{(1-x)_+} + 2283.011\: \delta (1-x) 
     - 722.137\: \ln^2 (1-x) \nonumber \\
  & & \mbox{} - 1236.264 - 332.254\: \ln x + 1.580\: 
     (\ln^4 x - 4\, \ln^3 x) \:\: .
\label{nsp0b}
\end{eqnarray}
The average $\frac{1}{2} \, [\, P^{(2)+}_{0,\, A}(x)\! +\! 
P^{(2)+}_{0,\, B} (x)\, ]$ represents our central result. 

The $N_f^1$-term $P_1^{(2)+}$ is the leading radiative correction to 
$P_1^{(1)+}(x)$, which is in turn only slightly more complicated
than the 1-loop non-singlet splitting function \cite{CFP}. Hence it is 
natural to adopt here the ansatz (\ref{pnlo}) employed for the 2-loop 
$N_f^0$-piece in our above illustration. The resulting eight 
approximations for $P_1^{(2)+}(x)$ are displayed for $x\! <\! 1$ in the 
left part of Fig.~4 (dashed curves). Their spread at large $x$ is 
similar to that obtained for $P_0^{(1)+}$ in Fig.~1. The leading 
large-$x$ coefficients fall into the range
\begin{equation}
     -190 \:\leq\: A_{1,1}^{(2)} \:\leq\: -180 \:\: .
\end{equation}
The uncertainty of the complete result for $P_{\rm NS} ^{(2)+}(x)$ 
is dominated by the spread of the above $N_f$-independent 
contribution, as estimated by the difference between Eqs.~(\ref{nsp0a}) 
and (\ref{nsp0b}). This is also true at small $x$, despite the fact 
that the band in Fig.~4 is presumably an underestimate in this region, 
as a possible term $\sim\ln^3 x$ has been disregarded. Hence it is 
sufficient, at the present stage, to keep only the $N_f^0$-contribution 
to the error band of $P_{\rm NS}^{(2)+}$ and to employ just one 
representative for $P_{1}^{(2)+}$. Our choice, an average of two 
typical results with and without a $\ln (1\! -\! x)$ term, reads
\begin{eqnarray} 
  P^{(2)+}_1(x) \: &=&
     \mbox{} - 184.4098\: \frac{1}{(1-x)_+} - 180.6971\: \delta (1-x) 
     - 98.5885\, \ln (1-x) 
     \nonumber \\
   & & \mbox{} + 205.7690\, x^2 + 6.1618\, + 5.0439 \ln^2 x
\label{nsp1}
\end{eqnarray}
and is also shown in the left part of Fig.~4 (solid curve).

As mentioned before the $N_f^2$-piece in Eq.~(\ref{nf_dep}) is exactly
known from ref.~\cite{Gra1}. After transformation to $x$-space, this
contribution reads 
\begin{eqnarray} 
  P^{(2)}_2(x) \: &=&
     \frac{1}{81} \bigg( \mbox{} - \frac{64}{(1-x)_+} 
     - [ 204 + 192\, \zeta (3) - 320\, \zeta (2) ] \,\delta (1-x)
     + 64
    \nonumber \\
  & & \mbox{}
     + \frac{x \ln x}{1-x}\, (96\,\ln x + 320)  
     + (1-x) (48\, \ln^2 x + 352 \ln x + 384) \bigg) \:\: ,
\end{eqnarray} 
where $\zeta (l)$ denotes Riemann's $\zeta$-function.

\begin{figure}[tbh]
\vspace*{-2mm}
\centerline{\epsfig{file=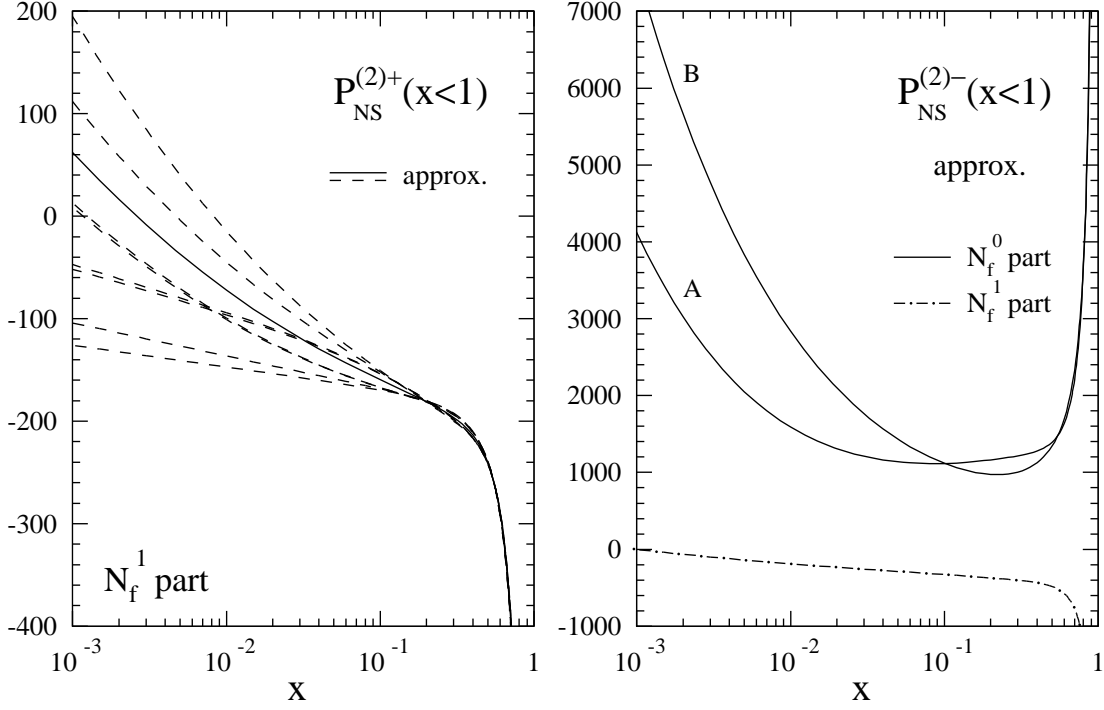,width=15cm,angle=0}}
\vspace*{-1mm}
\caption{Left: approximations to the $N_f^1$ part of $P^{(2)+}
 _{\rm NS}$, obtained from the five lowest even-integer moments using
 Eqs.~(\ref{ansatz}) and (\ref{pnlo}). Right: approximate results for 
 the $N_f^0$ and $N_f^1$ terms of the 3-loop splitting function
 $P^{(2)-}_{\rm NS}$.} 
\end{figure}

Finally we consider $P^{(2)-}_{\rm NS}(x)$. Here our treatment is 
inevitably more approximate. According to the expectations given at the 
beginning of this section, we take over the $1/(1-x)_+$ and $\delta 
(1-x)$ terms of the `+'-combinations in Eqs.~(\ref{nsp0a}), 
(\ref{nsp0b}), and (\ref{nsp1}). The remaining coefficients are (after 
inserting the appropriate $N_f^0$ leading small-$x$ piece \cite{BVns}) 
determined by the first, eighth and tenth moments of ref.~\cite{spfm}, 
assuming that the difference to $P^{(2)+}_{\rm NS}(x)$ is negligible 
for the latter two, entirely large-$x$ dominated quantities. 
The results are shown in the right half of Fig.~4. The uncertainty band 
for $P^{(2)-}_0(x)$ is about 50\% wider than that for $P^{(2)+}_0(x)$ 
around $x = 0.3$, reflecting the lack of precise knowledge of the
intermediate-$N$ moments, but smaller at small-$x$, as this region 
plays a much greater role for the first moment known here from 
Eq.~(2.6), than for the second moment in the `+'-case.
Our parametrizations spanning the present uncertainty are given by
\begin{eqnarray}
\label{nsm0a}
  P^{(2)-}_{0\, A}(x) \: &=& 
    1137.897 \frac{1}{(1-x)_+} + 1099.754\, \delta (1-x) 
    - 2954.684\, x^2 \nonumber \\
  & & \mbox{} - 143.709 - 2.761 \ln^2 x + 1.432\, \ln^4 x
  \\
  P^{(2)-}_{0,\, B}(x) \: &=& 
    1347.207\, \frac{1}{(1-x)_+} + 2283.011\, \delta (1-x) 
    - 722.238\, \ln^2 (1-x)
    \nonumber \\
  & & \mbox{} - 1234.756 -  327.479\, \ln x + 1.432\, (\ln^4 x - 4\,
    \ln^3 x) \:\: 
\label{nsm0b}
\end{eqnarray}
(also here the average represents the central result), supplemented by 
\begin{eqnarray} 
  P^{(2)-}_1(x) \: &=&
    \mbox{} - 184.4098\: \frac{1}{(1-x)_+} - 180.6971\: \delta (1-x) 
    - 98.5722\, \ln (1-x) 
    \nonumber \\
   & & \mbox{} + 205.3670\, x^2 + 6.5740\, + 3.5474 \ln^2 x \:\: .
\label{nsm1}
\end{eqnarray}
For the latter expression an average has been calculated in the same 
manner as for $P^{(2)+}_1$. 
%
%
\section{Numerical results}
%
\setcounter{equation}{0}
We are now ready to consider the numerical impact of the NNLO terms on 
the evolution of the non-singlet parton densities and structure 
functions. Before doing so, however, it is worthwhile to look at the
perturbative running of $\alpha_s$ underlying these considerations.
In the left part of Fig.~5 the $\alpha_s$-expansion (2.11) of the 
$\beta$-function is shown for $N_f = 4$ flavours. Besides the 
contributions of Eq.~(2.12) relevant for NNLO calculations, also the 
contribution $-\beta_3 \alpha_s^5 $ of ref.~\cite{beta3} has been 
included. If one uses the effect of this four-loop (N$^3$LO) term as an 
estimate of the residual error of the expansion, the resulting 
uncertainty amounts to 0.08\%, 0.35\%, 1.1\% and 2.5\% for $\alpha_s$ = 
0.12, 0.20, 0.30 and 0.40, respectively. The effects are somewhat 
larger (smaller) for $N_f\! =\! 3$ ($N_f\! =\! 5$). 
The consequences of this expansion on the scale dependence of 
$\alpha_s$ are illustrated in the right part of Fig.~5.  For this 
illustration we have used Eq.~(2.11) with $N_f = 3$ at $\mu_r \leq m_c
\! =\! 1.5$ GeV, $N_f = 4$ between $m_c$ and $m_b\! =\! 4.5$ GeV, and 
$N_f = 5$ for $\mu_r > m_b$, assuming that $\alpha_s(\mu_r^2)$ is 
continuous at these thresholds. If $\alpha_s$ is fixed to 0.115 at 
$\mu_r = M_Z$, then the four-loop effect reaches 0.1\% (1\%) only at 
$\mu_r^2 = 20 $ GeV$^2$ (1.5 GeV$^2$), respectively. Clearly the 
truncation of the series (2.11) after three terms does not introduce a 
significant theoretical uncertainty in the kinematic regime of 
deep-inelastic scattering. 

\begin{figure}[hbt]
\vspace*{2mm}
\centerline{\epsfig{file=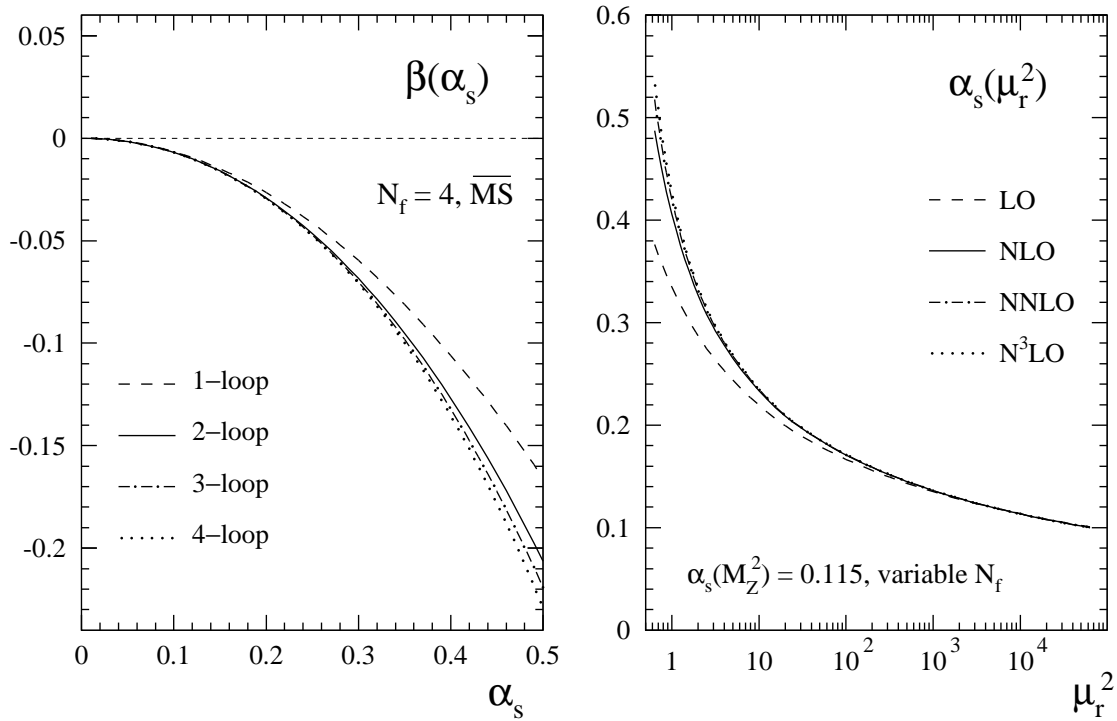,width=15cm,angle=0}}
\vspace{-1mm}
\caption{Left: The perturbative expansion of the QCD $\beta$-function 
 up to order $\alpha_s^5$, for four flavours in the $\overline
 {\mbox{MS}}$ renormalization scheme. Right: Illustration of the
 resulting scale dependence of $\alpha_s$, using a variable $N_f$ as 
 detailed in the text. $\mu_r^2$ is given in GeV$^2$.}
\end{figure}

For illustrations of the scale dependence of the parton densities and 
structure functions, initial distributions have to be chosen at some
reference scales, in the following denoted by $\mu_{f,0}^2$ and $Q_0^2$,
respectively, in Eqs.~(2.8) and (2.15). We will employ the function
\begin{equation}
  f \, = \, x^{0.5} (1-x)^3
\end{equation}

\vspace{-1cm}
\noindent
for all six quantities
\begin{equation}
  f \, = \, xq_{\rm NS}^{\pm}(x, \mu_{f,0}^2) \, , \:
  F_{2,\rm NS}^{\pm}(x, Q_0^2) \, \:\: \mbox{ and } \: 
  xF_{3,\rm NS}^{\pm}(x, Q_0^2) \:\: .
\end{equation}
Eq.~(5.1) represents a simple model shape which incorporates the most 
important features of non-singlet $x$-distributions of nucleons. The
same input is used in all cases, as this allows for a direct comparison 
of the effects of the various kernels in Eqs.~(2.10) and (2.16). The 
overall normalization of $f$ is irrelevant for the logarithmic 
derivatives considered below. Our initial scales are specified via
\begin{equation}
  \alpha_s (\mu_r^2\! = \!\mu_{f,0}^2) \, = \, 
  \alpha_s (\mu_r^2\! = \! Q_0^2) \, = \, 0.2 \:\: ,
\end{equation}
irrespective of the order of the expansion. For $\alpha_s (M_Z^2) = 
0.114 \ldots 0.120$ this choice corresponds to $\mu_{f,0}^2 = Q_0^2 
\,\simeq\, 25 \ldots 50$ GeV$^2$, a $Q^2$-region typical for 
fixed-target DIS. If not explicitly indicated otherwise, the results
will be given for $N_f = 4$ massless flavours.

\begin{figure}[bht]
\vspace*{1mm}
\centerline{\epsfig{file=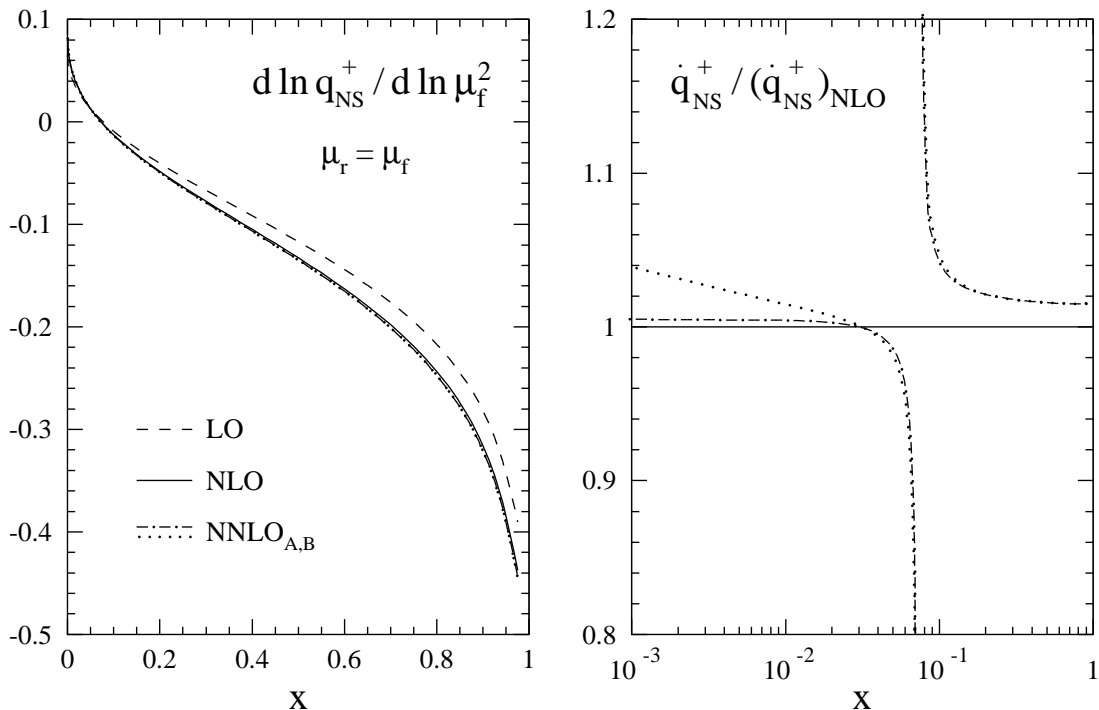,width=15cm,angle=0}}
\vspace{-1mm}
\caption{The perturbative expansion of the scale derivative, $\dot{q}
 _{\rm NS}^{\, +} \equiv d \ln q_{\rm NS}^{\, +}/ d\ln \mu_f^2$, for  
 a non-singlet `+'-combination of quark densities at $\mu_r = \mu_f$. 
 The initial conditions are as specified in Eqs.~(5.1)--(5.3).
 Here and in what follows the subscripts A and B indicate the 
 approximations for the 3-loop splitting functions derived in the 
 previous section.}
\end{figure}

The evolution of $q_{\rm NS}^{\, +}(x,\mu_f^2)$ is illustrated in 
Fig.~6 for the standard choice $\mu_r = \mu_f$ of the renormalization 
scale. In this case the perturbative expansion appears to be very well 
convergent: Except for the region around $x \simeq 0.07$ where the 
scale derivative is very small, the NNLO corrections for 
$\dot{q}_{\rm NS}^{\, +} \equiv d \ln q_{\rm NS}^{\, +}/ d\ln \mu_f^2$ 
are as small as about 2\%, while the NLO contributions typically 
amount to $10 \ldots 20 \%$. The residual uncertainty of the 3-loop 
splitting functions of Sect.~4 leads to a noticeable effect only for 
$x \lsim 0.02$, and even at $x \simeq 10^{-3}$ this effect does not 
exceed $\pm 2\%$ with respect to the central result $\frac{1}{2}
({\rm NNLO}_A + {\rm NNLO}_B)$ not shown in the figure. Over the full 
$x$-range the NNLO corrections are comparable to the dependence on
the number of active flavours: If $N_f$ is increased (decreased) to 
$N_f = 5$ ($N_f = 3$), $\dot{q}_{\rm NS}^{\, +}$ is decreased 
(increased) by about 2\%, respectively. 

\begin{figure}[bht]
\vspace*{2mm}
\centerline{\epsfig{file=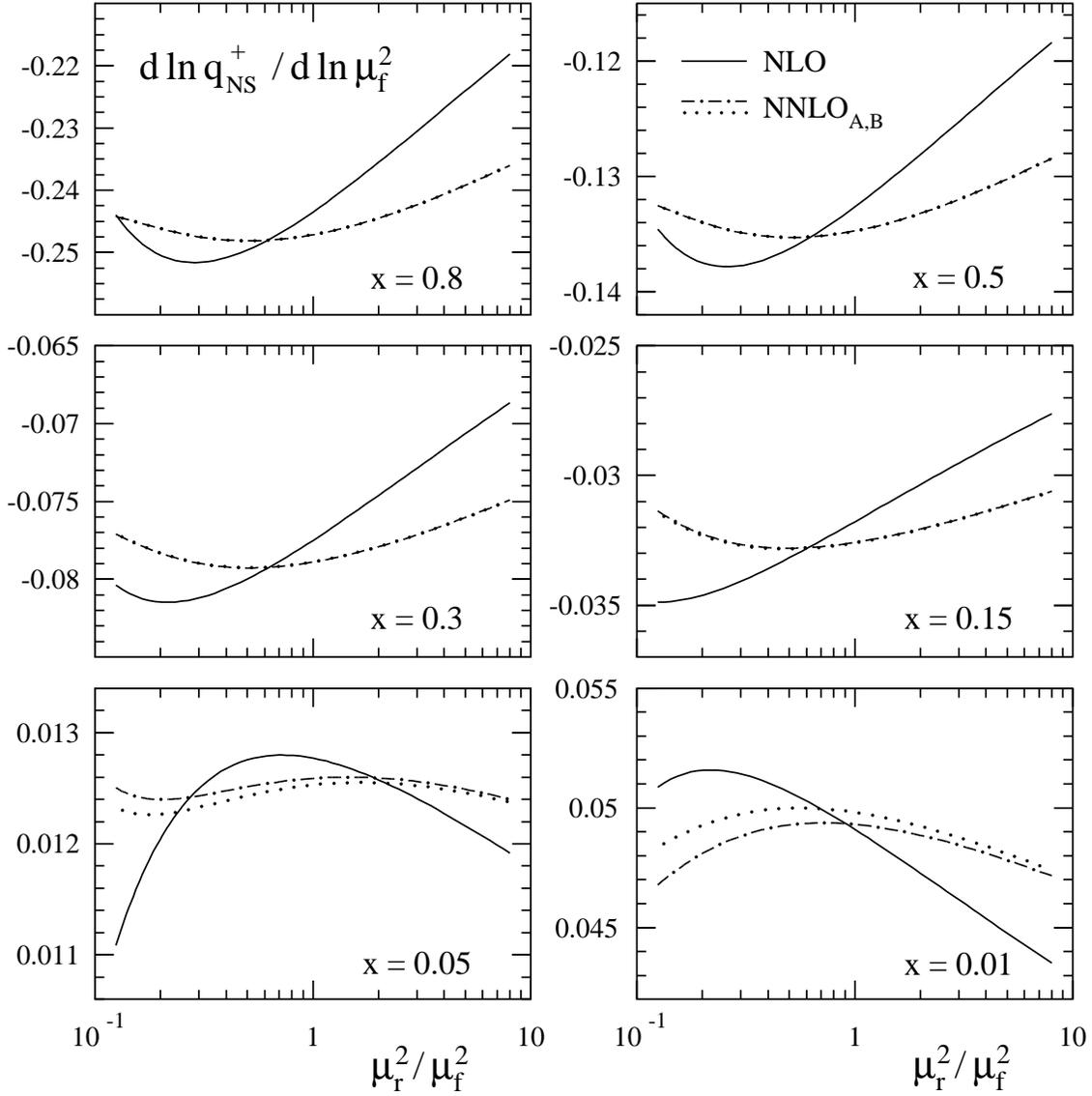,width=15cm,angle=0}}
\vspace*{-1mm}
\caption{The dependence of the NLO and NNLO predictions for 
 $d \ln q_{\rm NS}^{\, +}/ d\ln \mu_f^2$ on the renormalization scale 
 $\mu_r$ for six typical values of $x$.}
\label{figx}
\end{figure}

Another way to assess the reliability of perturbative calculations is 
to investigate the stability of the results under variations of the
renormalization scale $\mu_r$. In Fig.~7 the consequences of varying
$\mu_r$ over the rather wide range $\frac{1}{8}\mu_f^2 \leq \mu_r^2 
\leq 8 \mu_f^2$ are displayed for six representative values of $x$. 
The relative scale uncertainties of the average results, estimated by 
\begin{equation}
 \Delta \dot{q}_{\rm NS}^{\, +} \, \equiv \,  
 \frac{\max\, [ \dot{q}_{\rm NS}^{\, +}(x,\mu_r^2 = \frac{1}{4} 
 \mu_f^2 \ldots 4\mu_f^2)] - \min\, [\dot{q}_{\rm NS}^{\, +} (x, 
 \mu_r^2 = \frac{1}{4}\mu_f^2 \ldots 4 \mu_f^2)] } 
 { 2\, |\, {\rm average}\, [\dot{q}_{\rm NS}^{\, +}(x, \mu_r^2 = 
 \frac{1}{4}\mu_f^2 \ldots 4 \mu_f^2)]\, | }
\end{equation}
are shown in the left part of Fig.~8. Also this estimate leads to about 
2\% for the NNLO uncertainty, an improvement by more than a factor of 
three with respect to the corresponding NLO result. Even as low as 
$x \simeq 10^{-3}$ the NNLO calculation, despite its approximation 
uncertainty increasing towards small $x$, is superior to the NLO.

Finally the evolution of `$-$'-combinations $q_{\rm NS}^{\, -}$ is 
illustrated in the right part of Fig.~8. For $x > 0.1$ the difference 
to the `+'-case discussed so far is negligible at NLO as well as at 
NNLO.  At small $x$ the NLO predictions differ by up to 2\%. As 
expected from the discussion in Sect.~4, the residual uncertainty of 
the NNLO result is considerably 
less pronounced at small $x$ in the `$-$'-case, but somewhat larger for 
$ 0.01 \lsim x \lsim 0.1$.

\begin{figure}[bht]
\vspace*{2mm}
\centerline{\epsfig{file=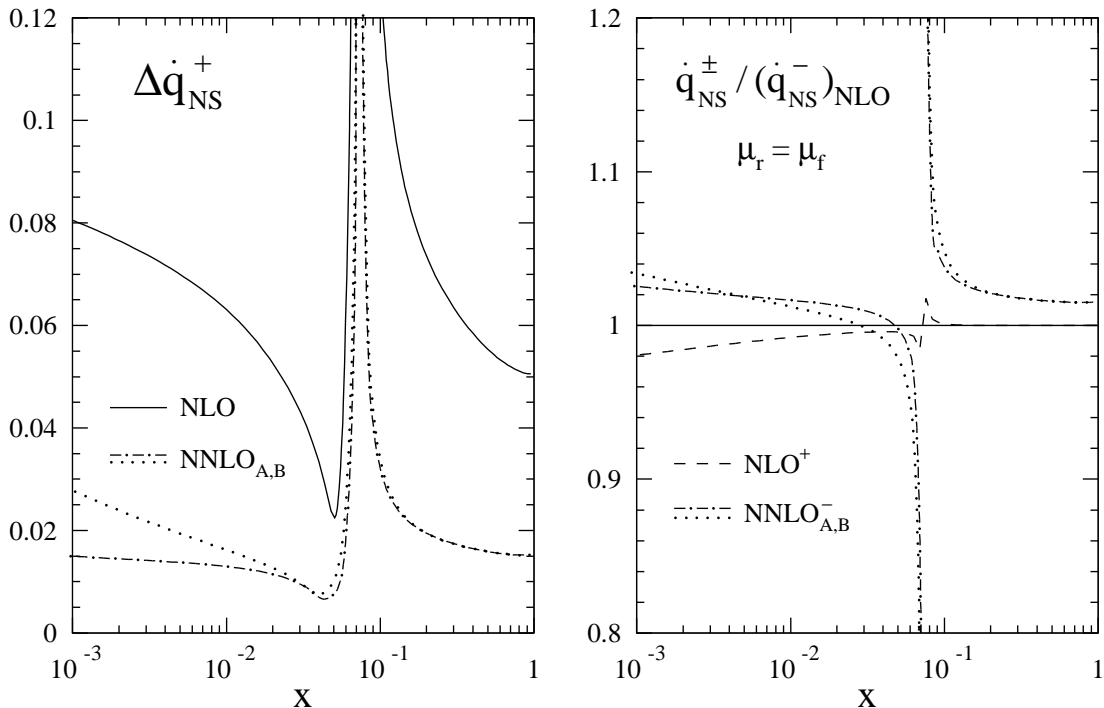,width=15cm,angle=0}}
\vspace*{-1mm}
\caption{Left: The renormalization scale uncertainty of the NLO and 
 NNLO predictions for the scale derivative of $q_{\rm NS}^+$, as 
 obtained from the quantity $\Delta \dot{q}_{\rm NS}^{\, +}$ defined in 
 Eq.~(5.4). Right: The NNLO effects on the evolution of $q_{\rm NS}
 ^{\, -}$ for the standard scale choice	$\mu_r = \mu_f$, together with
 a comparison of the NLO partonic NS$^+$ and NS$^-$ evolutions.}
\end{figure}

We now turn to the evolution of the non-singlet structure functions. 
The physical scale derivative $\dot{F}_{2,\rm NS}^{\, +} \equiv
d\ln F_{2,\rm NS}^{\, +}/ d\ln Q^2$ is shown in the left part of Fig.~9
for $\mu_r = Q$.
Besides the splitting functions $P_{\rm NS}^{(1,2)+}(x)$ the effect of 
which has been illustrated in Fig.~6, here also the coefficient 
functions $c_{2,\rm NS}^{(1)}(x)$ and $c_{2,\rm NS}^{(2)+}(x)$ enter 
the NLO and NNLO evolution kernels as detailed in Eq.~(2.16). These 
additional terms considerably increase the $Q^2$-dependence at large 
$x$, as can be seen by comparing Fig.~6 and Fig.~9. E.g., the NNLO 
corrections rise from 4\% at $x = 0.5$ to about 7, 11 and 21\% at 
$x = 0.65$, 0.8 and $x = 0.95$, respectively. The corresponding NLO 
contributions amount to 24, 30, 37 and 51\% of the LO results. 
Unlike for the parton densities, the NNLO corrections to the structure
functions are larger than the $N_f$-dependence at large $x$: If $N_f$ 
is increased (decreased) to $N_f = 5$ ($N_f = 3$), $\dot{F}_{2\rm NS}
^{\, +}$ is decreased (increased) between 3.5\% and 7\% for $0.5 \leq 
x \leq 0.95$,~respectively. 

\vspace*{-2mm}
The worse convergence of the expansion at large $x$ is due to the large 
soft-gluon contributions $[\ln^{k} (1\! -\! x)/(1\! -\! x)]_+$, 
$ k = 1, \ldots , 2l\! -\! 1$, to $c_{2,\rm NS}^{(l)\pm }(x)$ which 
are conjectured to be absent \cite{GLY} in the $\overline{\mbox{MS}}$ 
splitting functions $P_{\rm NS}^{(l)+}(x)$.  
Consequently, as shown in the right part of Fig.~9, keeping only the 
coefficient-function contributions in the $O(\alpha_s^3)$ term of 
Eq.~(2.16) yields a very good approximation at large $x$. In fact
$P_{\rm NS}^{(2)+}$ contributes less than 2\% to the total NNLO 
derivative $\dot{F}_{2,\rm NS}^{\, +}$ at $x > 0.2$. The residual 
uncertainty of $P_{\rm NS}^{(2)+}$, given by the difference 
NNLO$_A$ -- NNLO$_B$, is thus completely negligible in this region. 

\begin{figure}[bht]
\centerline{\epsfig{file=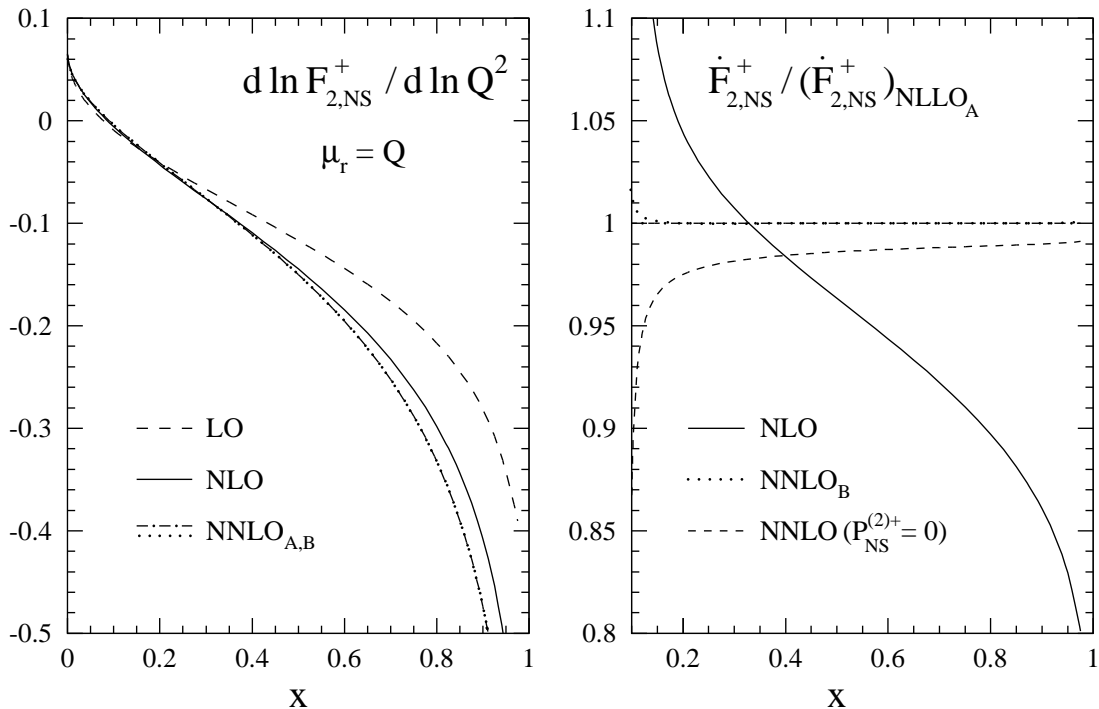,width=15cm,angle=0}}
\vspace*{-1mm}
\caption{The perturbative expansion of the scale derivative, 
 $\dot{F}_{2,\rm NS}^{\, +} \equiv d\ln F_{2,\rm NS}^{\, +}/ d\ln Q^2$, 
 for a non-singlet structure function at $\mu_r = \mu_f$. The initial 
 conditions are as specified in Eqs.~(5.1)--(5.3), the leading-order 
 curve is identical to that of Fig.~6. Also shown (right part) is the 
 effect of omitting the contribution from the 3-loop splitting 
 function $P_{\rm NS}^{(2)+}$.}
\end{figure}

The dependence of $\dot{F}_{2,\rm NS}^{\,+}$ on the renormalization 
scale $\mu_r$ is presented in Fig.~10 and Fig.~11 (left part), 
analogously to the partonic case (see Eq.~(5.4)) in Figs.~7 and~8 using
\begin{equation}
 \Delta \dot{F}_{2,\rm NS}^{\, +} \, \equiv \,  
 \frac{\max\, [ \dot{F}_{2,\rm NS}^{\, +}(x,\mu_r^2 = \frac{1}{4} Q^2 
 \ldots 4 Q^2)] - \min\, [\dot{F}_{2,\rm NS}^{\, +} (x, \mu_r^2 = 
 \frac{1}{4}Q^2 \ldots 4 Q^2)] } 
 { 2\, |\, {\rm average}\, [\dot{F}_{2,\rm NS}^{\, +}(x, \mu_r^2 = 
 \frac{1}{4} Q^2 \ldots 4 Q^2)]\, | } \:\: .
\end{equation}
The slower large-$x$ convergence of the $\alpha_s$ series for 
$\dot{F}_{2,\rm NS}^{\, +}$ is obvious from these results as well, 
e.g., no extremum close to $\mu_r = Q$ is obtained for $x = 0.8$. The 
NNLO uncertainties as estimated using Eq.~(5.5) read 3\%, 4.5\% and 7\% 
for $x = 0.5$, 0.65 and 0.8. The corresponding NLO results are 8.5\%, 
10.5\% and 12\%, respectively. The accuracy of the $Q^2$-slope 
predictions is thus improved by a factor $2 \ldots 3$ except for very 
large~$x$.

\begin{figure}[bht]
\vspace*{1mm}
\centerline{\epsfig{file=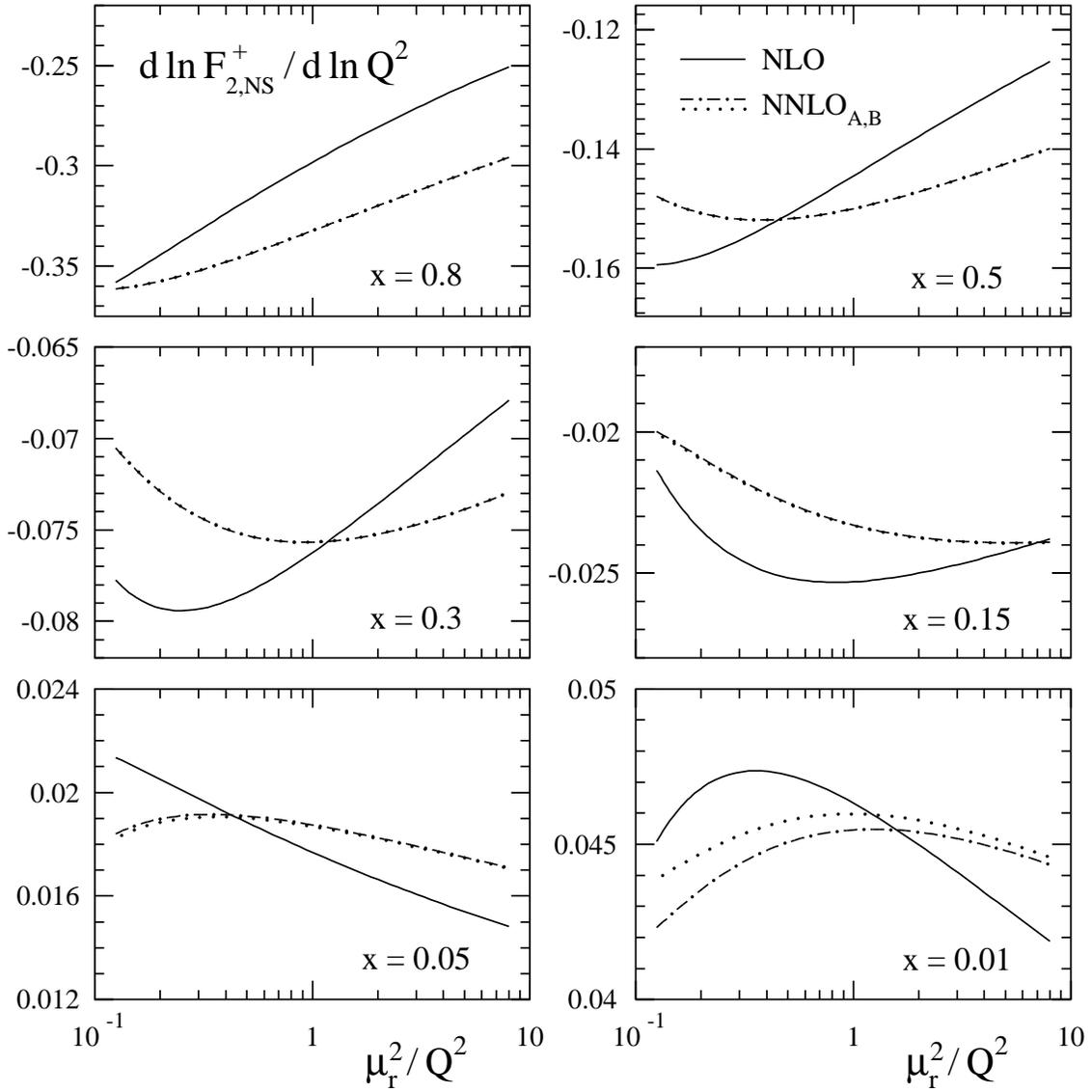,width=15cm,angle=0}}
\vspace*{-1mm}
\caption{The dependence of the NLO and NNLO predictions for 
 $d \ln F_{2,\rm NS}^{\, +}/ d\ln Q^2$ on the renormalization scale 
 $\mu_r$ for six typical values of $x$.} 
\end{figure}

\begin{figure}[thbp]
\vspace*{-2mm}
\centerline{\epsfig{file=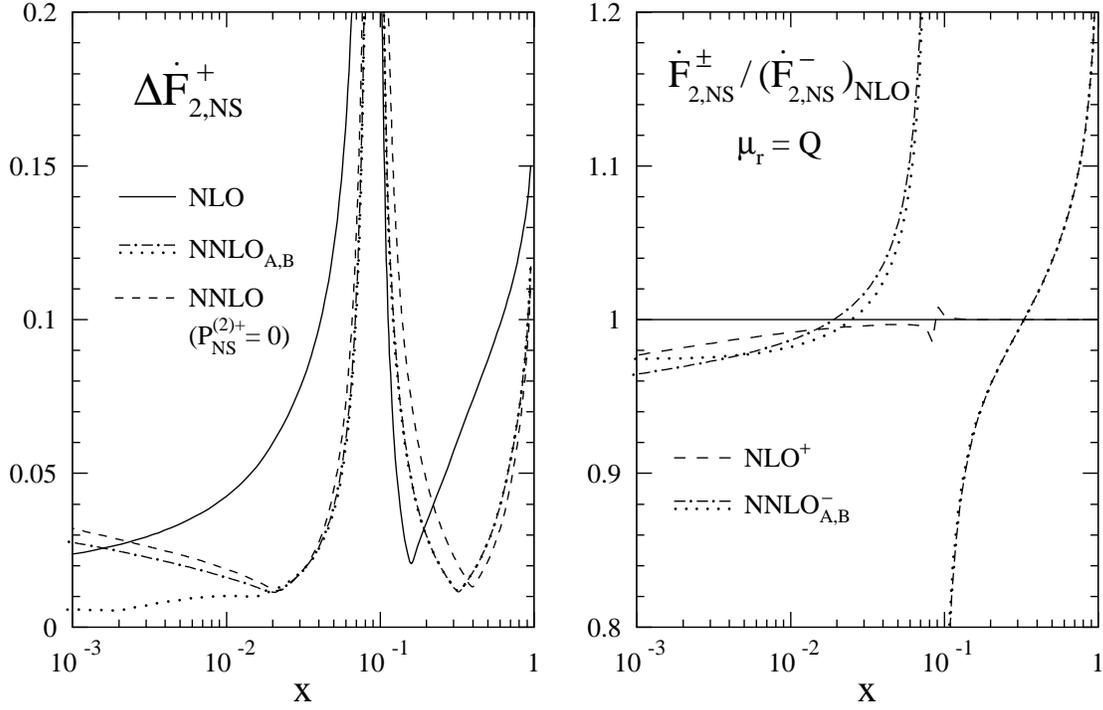,width=15cm,angle=0}}
\vspace*{-2mm}
\caption{Left: The $\mu_r$-uncertainty of the scale derivative of 
 $F_{2,\rm NS}^+$, as estimated by $\Delta \dot{F}_{2,\rm NS}^{\, +}$ 
 defined in Eq.~(5.5). Note that the absolute values of $\dot{F}_{2,\rm 
 NS}^{\, +}$  are very small for $0.05 < x < 0.15$. 
 Right: The NNLO effects on the evolution of $F_{2,\rm NS}^{\, -}$ for 
 $\mu_r = \mu_f$, together with a comparison of the NS$^+$ and NS$^-$ 
 evolutions for $F_{2,\rm NS}$ at NLO.}
\end{figure}

\begin{figure}[thbp]
\centerline{\epsfig{file=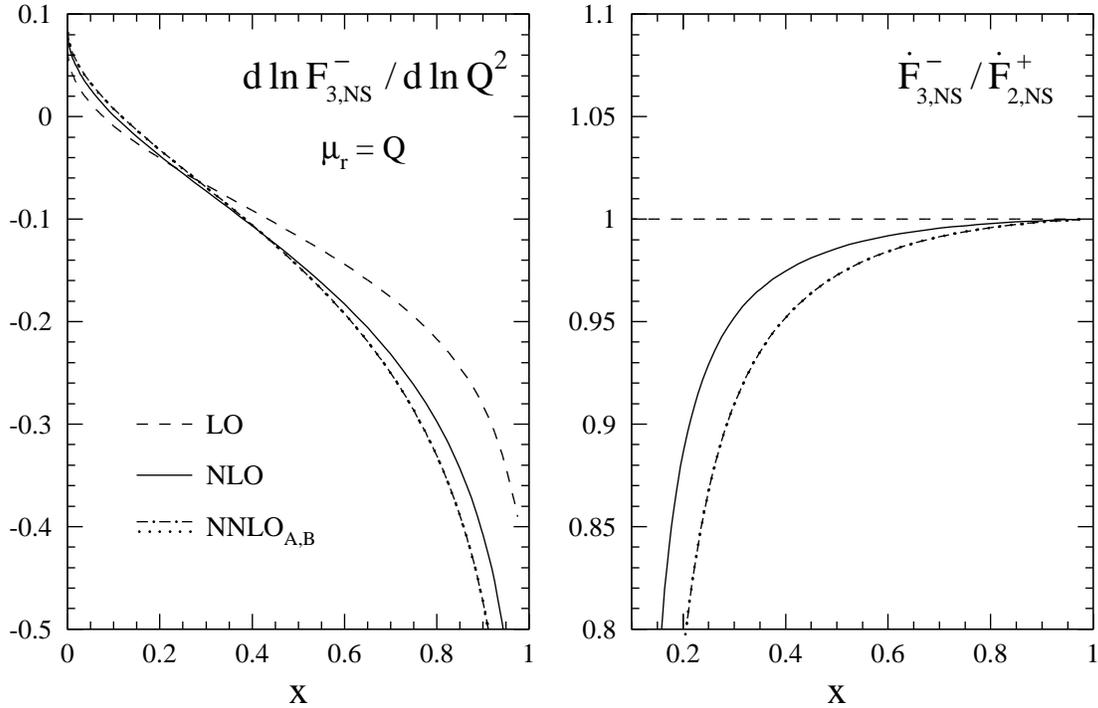,width=15cm,angle=0}}
\vspace*{-2mm}
\caption{The scale derivative $\dot{F}_{3,\rm NS}^{\, -} \equiv d\ln 
 F_{3,\rm NS}^{\, -}/ d\ln Q^2$ at $\mu_r = \mu_f$. The initial 
 conditions are as given in Eqs.~(5.1)--(5.3), the leading-order 
 curve is the same as in Figs.~6 and~9. Also shown (right part) 
 is the ratio of $\dot{F}_{3,\rm NS}^{\, -}$ to the corresponding 
 result for $F_{2,\rm NS}^{\, +}$.}
\end{figure}

As for the parton densities shown in Fig.~8, the evolution of 
$F_{2,\rm NS}^{\, -}$ illustrated in the right part of Fig.~11 is 
indistinguishable from that of $F_{2,\rm NS}^{\, +}$ at $x > 0.1$, 
while being better constrained at NNLO at very small $x$.  For 
$x \simeq 10^{-3}$, the $2.5\% \ldots 3.5\% $ positive effect of
$P_{\rm NS}^{(2)-}$ in Fig.~8 is overcompensated by the coefficient-%
function contributions. This effect also occurs for $\dot{F}_{2,\rm NS}
^{\, +}$ not displayed at small $x$. In both cases the NLO corrections 
are smaller than for $\dot{q}_{\rm NS}^{\pm}$, resulting in a better 
small-$x$ NLO renormalization-scale stability of 
$\dot{F}_{2,\rm NS}^{\,\pm}$ as can be seen by comparing the left
parts of Fig.~11 and Fig.~8.

The scaling violations of $F_{3,\rm NS}^{\, -}$ are presented in 
Fig.~12 for the medium- to large-$x$ region. Since the soft-gluon 
terms $[\ln^{k} (1\! -\! x)/(1\! -\! x)]_+$ are identical in 
$c_{2,\rm NS}^{\pm (l)}$ and $c_{3,\rm NS}^{\pm (l)}$, the results for 
$\dot{F}_{3,\rm NS}$ and $\dot{F}_{2,\rm NS}$ agree (for identical 
initial distributions as assumed here) as  $x \!\rightarrow \! 1$. 
However, the different regular terms lead to noticeable differences 
already at medium $x$, reaching 5\% and 10\% at $x \simeq 0.4$ and 0.3, 
respectively. At small $x$ the corrections are considerably larger for 
$\dot{F}_{3,\rm NS}$ than for $\dot{F}_{2,\rm NS}$, resulting in scale 
uncertainties of about 10\% at NLO and 4\% at NNLO for $10^{-3} \lsim x 
\lsim 10^{-2}$ for the former quantity.

Finally we turn to the determination of $\alpha_s$ from scaling 
violations of non-singlet structure functions. Here we address the 
uncertainties $\Delta \alpha_s$ which arise from the truncation of the 
perturbation series, confining ourselves to the region $x \gsim 0.25$ 
of considerable negative scale derivatives $\dot{F}_{2,NS}$. In this 
region the results for $\dot{F}_{3,\rm NS}$ are rather similar to those 
for $\dot{F}_{2,\rm NS}$ and need not to be considered separately. We 
also disregard the negligible large-$x$ differences between the scaling 
violations of $F_{2,\rm NS}^{\, +}$ and $F_{2,\rm NS}^{\, -}$ and 
between the NNLO$_A$ and NNLO$_B$ calculations.
Our procedure for estimating $\Delta \alpha_s$ is as follows: For each 
$x$ we determine those scales $\mu_{r,\rm min}$ and $\mu_{r,\rm max}$ 
which led to the minimal and maximal NLO and NNLO results for 
$|\dot{F}_{2,\rm NS}|$ used in Eq.~(5.5). The value of 
$\alpha_s(Q_0^2)$ is then adjusted to obtain, at these values of $x$ 
and $\mu_r$, the same results for $\dot{F}_{2,\rm NS}$ as found for 
$\mu_r = Q_0$ and $\alpha_s(Q_0^2) = 0.2$ (Fig.~9, left part). The 
latter standard-scale results thus play the role of the experimental 
results for $\dot{F}_{2,\rm NS}$ in determinations of $\Delta \alpha_s$ 
in data fits.

The resulting upper and lower limits for $\alpha_s(Q_0^2)$ are shown
in the left part in Fig.~13. Due to the increase of the higher-order
corrections towards large $x$ discussed above, the uncertainty $\Delta
\alpha_s$ rises with increasing $x$. As available experimental DIS 
results are restricted to $x \leq 0.85$ \cite{exp}, we choose a 
value $x \simeq 0.55$ for estimating the $x$-averaged uncertainties
given by the differences to the reference result $\alpha_s(Q_0^2) 
= 0.2$. This procedure yields
\begin{equation}
 \Delta \alpha_s (Q_0^2 \simeq 25 \ldots 50 \mbox{ GeV}^2)_{\rm NLO}  
 \:\:\: = 
 {\footnotesize \begin{array}{c} + \, 0.020 \\ -\, 0.012 \end{array}}
 \:\: ,
\end{equation}
\begin{equation}
 \Delta \alpha_s (Q_0^2 \simeq 25 \ldots 50 \mbox{ GeV}^2)_{\rm NNLO} 
 \, = 
 {\footnotesize \begin{array}{c} + \, 0.008 \\ -\, 0.004 \end{array}}
 \:\: .
\end{equation}
Often results and uncertainties for $\alpha_s$ from different processes
and observables are compared after evolution to a common reference 
scale, conventionally chosen as the Z-boson mass $M_Z$. 
Adopting $Q_0^2 = 30$ GeV$^2$ (and $N_f=5$ for $Q_0 \leq \mu_r \leq 
M_Z$) for definiteness, one obtains the error bands displayed in the 
right part of Fig.~13 and 
\begin{equation}
 \Delta \alpha_s (M_Z^2)_{\rm NLO} \: =
 {\footnotesize \begin{array}{l} + \: 0.006 \\ -\: 0.004 \end{array}}
 \: , \:\:\:\: 
 \Delta \alpha_s (M_Z^2)_{\rm NNLO} \: = 
 {\footnotesize \begin{array}{l} + \: 0.0025\\ -\: 0.0015\end{array}}
 \:\: .
\end{equation}
As expected from our previous discussions below Eq.~(5.5), the NNLO 
calculation reduces the theoretical uncertainty under consideration by 
a factor of about 2.5.

In a data analysis, also the NLO and NNLO central values for 
$\alpha_s(Q_0^2)$ will be different, since the NNLO scaling violations 
are stronger over most of the large-$x$ region as shown in Fig.~9. 
A simple estimate analogous to that for $\Delta \alpha_s$ yields
\begin{equation}
 \alpha_s(M_Z^2)_{\rm NNLO} - \alpha_s(M_Z^2)_{\rm NLO} \:\simeq\:
 -\, 0.002 \:\: .
\end{equation}
Due to the strong $x$-dependence of the NNLO/NLO ratio, this estimate
is less reliable than Eq.~(5.8), its uncertainty amounts to about 
$\pm 0.001$. Nevertheless it is interesting to note that Eq.~(5.9) 
agrees with the findings of ref.~\cite{KPS} from analyses of data on 
$F_3$. The 3-loop splitting function $P_{\rm NS}^{(2)}$ contribute only 
about $-0.0007$ to the shift (5.9) of the NNLO result.

\begin{figure}[bht]
\centerline{\epsfig{file=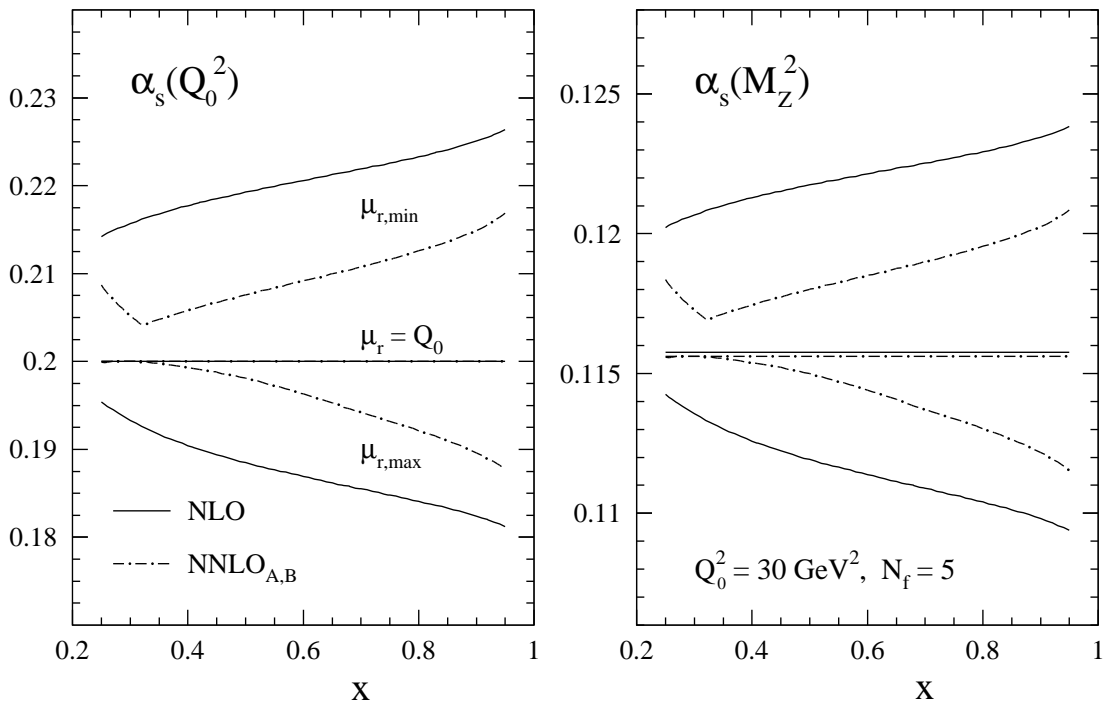,width=15cm,angle=0}}
\vspace*{2mm}
\caption{Right: The $x$-dependent theoretical uncertainty of the 
 determination of $\alpha_s$ from the scale derivative of 
 $F_{2,\rm NS}$ at $Q_0^2 = 25 \ldots 50 $ GeV$^2$, estimated by the 
 $\mu_r$-variation $\frac{1}{4} Q_0^2 \leq  \mu_r^2 \leq 4 Q_0^2$. 
 The scales leading to a maximal (minimal) $|\dot{F}_{2\rm NS}|$ are
 denoted by $\mu_{r,\rm max}$ ($\mu_{r,\rm min}$).
 Left: The resulting error band for $\alpha_s(M_Z^2)$ using $Q_0^2 = 
 30$ GeV$^2$ and $N_f = 5$.}
\end{figure}
%
%
\section{Summary}
%
%
We have investigated the effect of the NNLO perturbative QCD 
corrections on the scale dependence of flavour non-singlet quark
densities and structure functions.
For this purpose, and for application in further analyses, we have 
derived compact parametri\-zations of the corresponding three-loop 
splitting functions $P_{\rm NS}^{(2)\pm}$ and the two-loop coefficient 
functions $c_{a,\rm NS}^{(2)\pm}$, $a = 2, 3, L$. The latter
quantities are exactly known \cite{c2Sa,ZvN1,ZvN2}; their analytic
$x$-dependent expressions are however rather cumbersome and not readily 
transformed to moment space \cite{BlKu}. Our parametrizations of 
$c_{a,\rm NS}^{(2)\pm}(x)$ and their Mellin transforms thus provide
a convenient technical tool. They agree to the exact results up to a 
few permille or less over the full $x$-range, thus introducing a 
negligible error of well below 0.1\% after insertion into the 
perturbative expansions.
 
As only partial results are presently available for the three-loop 
splitting functions \cite{spfm,Gra1,BVns}, our parametrizations of 
$P_{\rm NS}^{(2)\pm}(x)$ serve the additional purpose of providing 
quantitative estimates of their $x$-dependent residual uncertainties. 
The function $P_{\rm NS}^{(2)+}(x)$, relevant to the evolution of 
flavour asymmetries like $u+\bar{u} - [d+\bar{d}]$, is well constrained 
at large $x$ by the lowest even-integer moments of refs.~\cite{spfm}, 
the spread reaching about $\pm 7\% $ at $x \simeq 0.3$. On the other 
hand $P_{\rm NS}^{(2)+}(x)$ is very weakly constrained for $x \lsim 
10^{-2}$ so far, despite the fact that the leading small-$x$ term is 
known \cite{BVns}. The quantity $P_{\rm NS}^{(2)-}(x)$, entering the 
evolution of the quark-antiquark differences, is somewhat better 
(worse) constrained at small $x$ (medium $x$), respectively, than 
$P_{\rm NS}^{(2)+}(x)$. 
As the splitting functions enter parton densities and structure 
functions only via convolutions with smooth non-perturbative initial
distributions, these `bare' uncertainties are very much reduced for 
physical quantities over the whole $x$-range. E.g., the spread of 
$P_{\rm NS}^{(2)+}$ leads to effects of less than $\pm 0.2\% $ at 
$x \gsim 0.2$ after convolution with typical nucleonic input shapes. In 
this region the present uncertainties of $P_{\rm NS}^{(2)\pm}$ are thus 
rendered absolutely negligible, leading to effects even below 0.01\% 
after insertion into the perturbation series. Their impact becomes 
significant only for $x\lsim 10^{-2}$, without seriously impairing the
NNLO calculations even down to $x \simeq 10^{-3}$.
 
The perturbative expansion for the scale dependence $d \ln q_{\rm NS}
^{\,\pm}(x,\mu_f^2)/ d\ln \mu_f^2$ of the non-singlet combinations of 
quark densities appears to be very well convergent. For $\alpha_s = 
0.2$, corresponding to scales of about 25--50 GeV$^2$, the NNLO effects 
of $P_{\rm NS}^{(2)\pm}$ are on the level of 2\% rather uniformly in 
$x$. This result is to be compared to the NLO corrections which amount 
to 10--20\%. Also the variation of the renormalization scale leads to 
effects of about $\pm 2\%$ at NNLO. Corrections of this size are 
comparable to the dependence of the predictions on the number of quark 
flavours, rendering a proper treatment of charm effects \cite{chrm} 
rather important even for large-$x$ non-singlet quantities. 

Especially at $x > 0.5$, the higher-order corrections are much larger
for the scale derivative $d\ln F_{a,\rm NS}^{\,\pm}(x,Q^2)/ d\ln Q^2$,
$a = 2, 3$, of the non-singlet structure functions. This enhancement is
an effect of the coefficient functions containing large
$[\ln^{k} (1\! -\! x)/(1\! -\! x)]_+$ soft-gluon terms, which are 
conjectured to be absent in the $\overline{\mbox{MS}}$ splitting 
functions \cite{GLY}. E.g., the NLO and NNLO effects reach about 37\% 
and 11\% of the respective lower-order results at $x = 0.8$ for 
$\alpha_s = 0.2$ and four flavours. The NNLO calculations thus 
represent a distinct improvement, reducing also the renormalization-%
scale dependence of the predictions by a factor of two to three, e.g., 
to about $\pm 7\%$ at $x = 0.8$. 
Accordingly the inclusion of the NNLO corrections into fits of data on 
non-singlet scaling violations is expected to yield, besides a slight 
lowering of the central values for $\alpha_s(M_Z^2)$ by roughly 0.002, 
a considerable reduction of the (so far dominant) theoretical error due 
to the truncation of the perturbation series,
$$  
 \Delta \alpha_s (M_Z^2)_{\rm NNLO} \: = 
 {\footnotesize \left. \begin{array}{l} + \: 0.0025\! \\ -\: 0.0015 
 \end{array} \right| }_{\, {\mbox {\small $\mu $}}_
 {\mbox{\scriptsize $r$}}} \:\: .
$$
These estimates are compatible with the results of the fits of 
$F_{3,\rm NS}$-data performed in ref.~\cite{KPS}, where an 
alternative, integer-moment based approach to the calculation of the 
scaling violations has been pursued.

{\sc Fortran} subroutines of our parametrizations of $c_{a,\rm NS}
^{(2)\pm}(x)$, $a = 2, 3, L$, and $P_{\rm NS}^{(2)\pm}(x)$ can by 
obtained via email to neerven@lorentz.leidenuniv.nl or 
avogt@lorentz.leidenuniv.nl.
%
%
\section*{Acknowledgment}
%
%
This work has been supported by the European Community TMR research
network `Quantum Chromodynamics and the Deep Structure of Elementary 
Particles' under contract No.~FMRX--CT98--0194.
%
%
%
\newpage
\section*{Appendix: Third-order quantities in Mellin-{\boldmath $N$} 
          space}
%
\setcounter{equation}{0}
\renewcommand{\theequation}{A.\arabic{equation}}
The Mellin transforms of the approximate NNLO expressions of Sect.~3 
and Sect.~4 are given in terms of the integer-$N$ sums $S_l(N)$ and 
their analytic continuations
\begin{equation}
  S_{l} \equiv S_{l}(N) = \sum_{k=1}^N \frac{1}{k^l}
  = \zeta (l) - \frac{(-1)^l}{(l-1)!} \,\psi^{(l-1)}(N\! +\! 1)
  \:\: .
\end{equation}
Here $\zeta (1) $ stands for the Euler--Mascheroni constant, and 
$\zeta (l\! >\! 1)$ for Riemann's $\zeta$-func\-tion. The $l\,$th 
logarithmic derivative $\psi^{(l-1)}$ of the $\Gamma$-function can be 
readily evaluated using the asymptotic expansion for ${\rm Re}\, N>10$ 
together with the functional equation.
 
Due to the simplicity of our parametrizations for the three-loop
splitting functions, only the most simple Mellin transforms occur for 
these quantities. Therefore we are able to dispense with details here.
The Mellin-$N$ dependence of the exactly known $N_F^2$-piece can be 
found in ref.~\cite{Gra1}.

The moments the non-singlet `+'-coefficient function (3.2) entering
$F_2^{\, {\rm e.m.}}$ are given by  
\begin{eqnarray}
  c_{2,{\rm NS}}^{(2)+}(N) \!
   &=\! & \mbox{}+ 3.55555\, \Big( 6\, S_4 + 8\, S_3 S_1 + 3\, S_2^2 
       + 6\, S_2 S_1^2 + S_1^4 \Big)
       \nonumber \\
   & & \mbox{}+ 20.4444 \,\Big( 2\, S_3 + 3\, S_2 S_1 + S_1^3 \Big)
       - 15.5525\, \Big( S_2 + S_1^2 \Big) - 188.64\, S_1
       \nonumber \\
   & & \mbox{}+ \frac{165.356}{N} S_3
       - \bigg( \frac{15.38}{N^2} - \frac{9.7467}{N} \bigg) S_2 
       + \frac{358.503}{N} S_2 S_1 + \frac{2.9678}{N} S_1^3
       \nonumber \\
   & & \mbox{}+ \bigg( \frac{174.8}{N^2} + \frac{9.7467}{N} \bigg) S_1^2
       - \bigg( \frac{190.18}{N^3} - \frac{116.734}{N} \bigg) S_1 
       \nonumber \\
   & & \mbox{}+ \frac{17.01}{N^4} - \frac{34.16}{N^3} 
       + \frac{306.849}{N^2} - \frac{72.5824}{N} - \frac{1008}{N+1} 
       - 338.044 
       \\
   & & \hspace*{-7mm}\mbox{}+ N_f\: \bigg\{ 
       - 0.59259\, \Big( 2\, S_3 + 3\, S_2 S_1 + S_1^3 \Big)
       - 4.2963\, \Big( S_2 + S_1^2 \Big) - 6.3489\, S_1
       \nonumber \\
   & & \mbox{} - \frac{6.072}{N} S_3 - \bigg( \frac{6.072}{N^2} - 
       \frac{18.0408}{N} \bigg) S_2
        - \bigg( \frac{6.072}{N^3} - \frac{17.97}{N^2} 
        + \frac{14.3574}{N} \bigg) S_1 
       \nonumber \\
   & & \mbox{} + \frac{0.07078}{N} S_1^2 
    + \frac{4.488}{N^3} + \frac{4.21808}{N^2} 
       - \frac{21.6028}{N} - \frac{37.91}{N+1} + 46.8406 \bigg\}
       \:\: . \nonumber
\end{eqnarray}    
For the charged current `$-$'-combination the third to fifth
line of this result are, according to Eq.~(3.3), replaced by
\begin{eqnarray}
   & & \mbox{}+ \frac{229.916}{N} S_3
       + \bigg( \frac{31.58}{N^2} + \frac{9.7467}{N} \bigg) S_2 
       + \frac{393.703}{N} S_2 S_1 + \frac{2.9678}{N} S_1^3
       \nonumber \\
  & & \mbox{}+ \bigg( \frac{192.4}{N^2} + \frac{9.7467}{N} \bigg) S_1^2 
       - \bigg( \frac{160.82}{N^3} - \frac{61.1321}{N} \bigg) S_1 
       \\
   & & \mbox{}+ \frac{22.488}{N^4} - \frac{39.12}{N^3} 
       + \frac{265.774}{N^2} - \frac{164.777}{N} - \frac{1010}{N+1} 
       - 337.992 \:\: .
       \nonumber
\end{eqnarray}
The first two lines and the sixth line of Eq.~(A.2) stem from 
the universal +-distribution parts of Eq.~(3.2). They are exact up to
a truncation of the numerical factors \cite{ZvN1}. 
 
\noindent
The corresponding $N$-space results for the coefficient functions (3.4) 
and (3.5) for $F_L$ read
\begin{eqnarray}
  c_{L,{\rm NS}}^{(2)+}(N) \!
   &=\!& \mbox{} - \frac{136.88}{N} S_2 + \frac{13.62}{N} S_1^2 + 
         \bigg( \frac{55.79}{N} - \frac{150.5}{N^2} \bigg) S_1
         - \frac{0.062}{N^3}  \nonumber \\
   & & \mbox{} + \frac{14.85}{N^2} + \frac{207.153}{N}
        + \frac{53.12}{(N+1)^3} + \frac{97.48}{N+1} - 0.164 \\
   & & \hspace*{-6mm}\mbox{}+ N_f\: \frac{16}{27} \: \bigg\{ 
       - \frac{6}{N+1} S_1 + \frac{6}{N} + \frac{6}{(N+1)^2} 
       - \frac{25}{N+1} \bigg\} \nonumber 
\end{eqnarray}
and 
\begin{eqnarray}
  c_{L,{\rm NS}}^{(2)-}(N) \!
   &=\!& \mbox{} - \frac{128.4}{N} S_2 + \frac{13.30}{N} S_1^2 + 
         \bigg( \frac{59.12}{N} - \frac{141.7}{N^2} \bigg) S_1
         - \frac{0.086}{N^3} \nonumber \\
   & & \mbox{} + \frac{22.21}{N^2} + \frac{180.818}{N} 
        + \frac{46.58}{(N+1)^3} + \frac{100.8}{N+1} - 0.150 \\
   & & \hspace*{-6mm}\mbox{}+ N_f\: \frac{16}{27} \: \bigg\{ 
       - \frac{6}{N+1} S_1 + \frac{6}{N} + \frac{6}{(N+1)^2} 
       - \frac{25}{N+1} \bigg\} \nonumber \:\: .
\end{eqnarray}
Here the $N_f$ parts represent an exact result \cite{c2Sa}.

\noindent
The `$-$'-coefficient function (3.6) for $F_3$, occurring in the 
$\nu \! + \!  \bar{\nu} $ sum, leads to 
\begin{eqnarray}
  c_{3,\rm NS}^{(2)-}(N) \!
   &=\! & \mbox{}+ 3.55555\, \Big( 6\, S_4 + 8\, S_3 S_1 + 3\, S_2^2 
       + 6\, S_1^2 S_2 + S_1^4 \Big)
       \nonumber \\
   & & \mbox{}+ 20.4444 \,\Big( 2\, S_3 + 3\, S_2 S_1 + S_1^3 \Big)
       - 15.5525\, \Big( S_2 + S_1^2 \Big) - 188.64\, S_1
       \nonumber \\
   & & \mbox{}+ \frac{297.756}{N} S_3
       + \bigg( \frac{147.9}{N^2} + \frac{33.2767}{N} \bigg) S_2 
       + \frac{298.733}{N} S_2 S_1 + \frac{0.9778}{N} S_1^3
       \nonumber \\
   & & \mbox{}+ \bigg( \frac{147.9}{N^2} + \frac{33.2767}{N} \bigg) 
       S_1^2 - \frac{45.8683}{N} S_1 
       \nonumber \\
   & & \mbox{}+ \frac{23.532}{N^4} - \frac{66.62}{N^3} 
       + \frac{67.6}{N^2} - \frac{373.029}{N} - \frac{576.8}{N+1} 
       - 338.625
       \\
   & & \hspace*{-7mm}\mbox{}+ N_f\: \bigg\{ 
       - 0.59259\, \Big( 2\, S_3 + 3\, S_2 S_1 + S_1^3 \Big)
       - 4.2963\, \Big( S_2 + S_1^2 \Big) - 6.3489\, S_1
       \nonumber \\
   & & \mbox{} - \frac{0.042}{N} \Big (2\, S_3 + 3\, S_2 S_1 + S_1^3
       \Big)  + \frac{0.96978}{N} S_1^2
        + \bigg( \frac{9.684}{N^2} 
        - \frac{16.4074}{N} \bigg) S_1 
       \nonumber \\
   & & \mbox{} 
    +  \frac{10.6538}{N} S_2 + \frac{4.414}{N^3} - \frac{8.683}{N^2} 
       - \frac{15.9177}{N} - \frac{14.97}{N+1} + 46.856 \bigg\} \:\: .
       \nonumber
\end{eqnarray}    
For the `+'-combination of Eq.~(3.7) entering $F_3^{\nu N} - 
F_3^{\bar{\nu} N}$ one has to replace the third to fifth line of the 
above result by
\begin{eqnarray}
   & & \mbox{}+ \frac{186.816}{N} S_3
       + \bigg( \frac{92.43}{N^2} + \frac{ 33.2767}{N} \bigg) S_2 
       + \frac{187.793}{N} S_2 S_1 + \frac{0.9778}{N} S_1^3
       \nonumber \\
   & & \mbox{}+ \bigg( \frac{92.43}{N^2} + \frac{33.2767}{N} \bigg)  
       S_1^2 + \frac{123.121}{N}  S_1 
       \\
   & & \mbox{}+ \frac{18.294}{N^4} - \frac{60.28}{N^3} 
       + \frac{79.14}{N^2} - \frac{ 276.473}{N} - \frac{467.2}{N+1} 
       - 338.681 \:\: .
       \nonumber
\end{eqnarray}
%
%

%

\newpage
\begin{thebibliography}{99}
%
%
\bibitem{SLAC}  D.H. Coward et al., Phys.\ Rev.\ Lett.\ {\bf 20} (1968) 
                292; \\
                E.D. Bloom et al., Phys.\ Rev.\ Lett.\ {\bf 23} (1969) 
                930; \\
                H. Breitenbach et al., Phys.\ Rev.\ Lett.\ {\bf 23} 
                (1969) 935
\bibitem{exp}   C. Caso et al., Particle Data Group, Eur.\ Phys.\ J. 
                {\bf C3} (1998) 1, and references therein
\bibitem{FP82}  W.\ Furmanski and R.\ Petronzio, Z. Phys.\ {\bf C11} 
                (1982) 293, and references therein
\bibitem{beta2} O.V. Tarasov, A.A. Vladimirov, and A.Yu.\ Zharkov,
                Phys.\ Lett.\ {\bf B93} (1980) 429; \\
                S.A. Larin and J.A.M. Vermaseren, Phys.\ Lett.\ 
                {\bf B303} (1993) 334
\bibitem{c2Sa}  J. Sanchez Guillen et al., Nucl.\ Phys.\ {\bf B353}
                (1991) 337
\bibitem{ZvN1}  E.B. Zijlstra and W.L. van Neerven, Phys.\ Lett.\ 
                {\bf B272} (1991) 127, ibid.\ {\bf B273} (1991) 476,
                ibid.\ {\bf B297} (1992) 377
\bibitem{ZvN2}  E.B. Zijlstra and W.L. van Neerven, Nucl.\ Phys.\ 
                {\bf B383} (1992) 525; \\ 
                E.B. Zijlstra, thesis, Leiden University 1993 
\bibitem{spfm}  S.A. Larin, T. van Ritbergen, and J.A.M. Vermaseren,
                Nucl.\ Phys.\ {\bf B427} (1994) 41; \\
                S.A. Larin, P. Nogueira, T. van Ritbergen, and J.A.M. 
                Vermaseren, Nucl.\ Phys.\ {\bf B492} (1997) 338, \\
                T. van Ritbergen, thesis, Amsterdam University 1996
\bibitem{c2DY}  R. Hamberg, W.L. van Neerven and T. Matsuura, Nucl.\
                Phys.\ {\bf B359} (1991) 343; \\ 
                R. Hamberg, thesis, Leiden University 1991; \\
                W.L. van Neerven and E.B. Zijlstra, Nucl.\ Phys.\ 
                {\bf B382} (1992) 11
\bibitem{MRS}   A.D. Martin, R.G. Roberts and W.J. Stirling, Phys.\ 
                Lett.\ {\bf B387} (1996) 419; \\
                A.D. Martin, R.G. Roberts, W.J. Stirling and R.S. 
                Thorne, Eur.\ Phys.\ J. {\bf C4} (1998) 463 
\bibitem{CTEQ}  H.L. Lai et al., CTEQ Collab., Phys.~Rev.\ {\bf D55} 
                (1997) 1280; Michigan State University preprint 
                MSU-HEP-903100 ({\tt hep-ph/9903282})
\bibitem{GRV}   M.\ Gl\"uck, E.\ Reya and A.\ Vogt, Z.\ Phys.\ 
                {\bf C67} (1995) 433; Eur.\ Phys.\ J.\ {\bf C5} 
                (1998)~461 
\bibitem{KPS}   A.L. Kataev, A.V. Kotikov, G. Parente and A.V. Sidorov, 
                Phys.\ Lett.\ {\bf B388} (1996) 179; ibid.\ {\bf B417} 
                (1998) 374; \\ 
                A.L. Kataev, G. Parente, A.V. Sidorov, ICTP preprint
                IC/99/51 ({\tt hep-ph/9905310})
\bibitem{SY99}  J. Santiago and F.J. Yndurain, Madrid University
                preprint FTUAM-99-8 ({\tt hep-ph/ 9904344})
\bibitem{Gra1}  J. A. Gracey, Phys.\ Lett.\ {\bf B322} (1994) 141
\bibitem{Gra2}  J.F. Bennett and J.A. Gracey, Nucl.\ Phys.\ {\bf B517}
                (1998) 241
\bibitem{CH94}  S. Catani and F. Hautmann, Nucl.\ Phys.\ {\bf B427} 
                (1994) 475 
\bibitem{BVns}   J. B\"umlein and A. Vogt, Phys.\ Lett.\ {\bf B370} 
                (1996) 149 
\bibitem{BNRV}  J. Bl\"umlein and A. Vogt, Phys.\ Rev.\ {\bf D58} 
                (1998) 014020; \\
                J. Bl\"umlein, V. Ravindran, W.L. van Neerven and 
                A. Vogt, Proceedings of DIS~98, Brussels, April 1998,
                eds.\ Gh. Coremans and R. Roosen (World Scientific 
                1998), p.~211 ({\tt hep-ph/9806368})  
\bibitem{cmom}  M. Diemoz, F. Ferroni, E. Longo and G. Martinelli, 
                Z.\ Phys.\ {\bf C39} (1988) 21; \\
                M. Gl\"uck, E. Reya and A. Vogt, Z.\ Phys.\ {\bf C48} 
                (1990) 471; \\
                Ch.\ Berger, D. Graudenz, M. Hampel and A. Vogt, Z.\
                Phys.\ {\bf C70} (1996) 77
\bibitem{BlKu}  J. Bl\"umlein and S. Kurth, DESY preprint 97-160 
                ({\tt hep-ph/9708388}); Phys.\ Rev.\ {\bf D60} (1999) 
                014018
\bibitem{KL83}  J. Kirschner and L.N. Lipatov, Nucl.\ Phys.\ {\bf B213}
                (1983) 122
\bibitem{CFP}   G. Curci, W. Furmanski and R. Petronzio, Nucl.\ Phys.\
                {\bf B175} (1980) 27
\bibitem{GLY}   A. Gonzales-Arroyo, C. Lopez and F.J. Yndurain, Nucl.\ 
                Phys.\ {\bf B126} (1979) 161
\bibitem{BRV}   J. Bl\"umlein, S. Riemersma and A. Vogt, Nucl.\ Phys.\ 
                (Proc.\ Suppl.) {\bf 51C} (1996) 30 ({\tt 
                hep-ph/9608470}) 
\bibitem{beta3} T. van Ritbergen, J.A.M. Vermaseren and S.A. Larin, 
                Phys.\ Lett.\ {\bf B400} (1997)~379
\bibitem{chrm}  E. Laenen, S. Riemersma, J. Smith, W.L. van Neerven,
                Nucl.\ Phys.\ {\bf B392} (1993)~162; \\
                M. Buza, Y. Matiounine, J. Smith and W.L. van Neerven, 
                Eur.\ Phys.\ J.\ {\bf C1} (1998) 301, and references
                therein 
%
\end{thebibliography}
\end{document}